\documentclass{article}
\usepackage[margin=3.7cm]{geometry}
\usepackage{amsmath,amssymb,amsbsy,graphicx,float, color}
\newcommand{\nl}{\nonumber\\&\phantom{=}}
\newcommand{\RR}{{\mathbb{R}}}
\newcommand{\CC}{{\mathbb{C}}}
\newcommand{\pa}{\partial}
\newcommand{\ii}{{\rm i}}
\newcommand{\dd}{{\rm d}}
\newcommand{\tr}{\operatorname{tr}}
\newcommand{\bx}{\boldsymbol{x}}
\newcommand{\by}{\boldsymbol{y}}
\newcommand{\bS}{\boldsymbol{S}}
\newcommand{\bP}{\boldsymbol{P}}
\newcommand{\bL}{\boldsymbol{L}}
\newcommand{\brho}{\boldsymbol{\rho}}
\newcommand{\bomega}{\boldsymbol{\omega}}
\newcommand{\bsigma}{\boldsymbol{\sigma}}
\newcommand{\tautau}{\boldsymbol{\tau}_1\boldsymbol{\tau}_2}
\newcommand{\sigsig}{\boldsymbol{\sigma}_1\boldsymbol{\sigma}_2}
\newcommand{\Ls}{\boldsymbol{L}\cdot\boldsymbol{\sigma}}
\usepackage{hyperref}

\hypersetup{
	colorlinks = false,
}

\title{
	\vskip -50pt
	\bf{Nucleon-nucleon potential from instanton holonomies}}
\author{Chris Halcrow$^\ast$
	 and Derek Harland$^\dagger$
	\bigskip
	\\$^\ast$Department of Physics, KTH-Royal Institute of Technology, Sweden
	\\email address: chalcrow@kth.se
	\bigskip
	\\$^\dagger$School of Mathematics, University of Leeds, UK
	\\email address: d.g.harland@leeds.ac.uk
}
\date{9th August 2022}

\begin{document}

\maketitle
	
\abstract
We derive the nucleon-nucleon interaction from the Skyrme model using the instanton and product approximations to skyrmion dynamics. In doing so, we also calculate the classical potential and metric for skyrmion dynamics in each of the approximations. This is the first time they have been compared in detail and the results show major disagreements between the approximations. We derive the eight low energy nucleon-nucleon interaction potentials and compare them with the Paris model. For the instanton approximation we find strong negative isoscalar and isovector spin-orbit potentials, matching phenomenological models and our geometric intuition. Results for the other potentials are mixed, in part due to the zero pion mass limit used in this approximation.
	
\section{Introduction}

Understanding the nucleon-nucleon interaction is one of the fundamental problems in nuclear physics.  Most models treat nucleons as point particles with spin and isospin degrees of freedom, and consider the most general system consistent with the underlying symmetries of QCD. In the most basic set-up the quantum nucleon-nucleon hamiltonian consists of eight terms, each arising with an effective potential depending on the nucleon separation. The potentials are fitted using theoretical and experimental inputs, in various different ways \cite{Lacombe:1980dr, MACHLEIDT19871, PhysRevC.51.38}.

Skyrme proposed an alternative theory of nuclei in the 1960s where nucleons arise as localised collective excitations of pions, which owe their existence and stability to the topology of the system. Now called skyrmions, these nonlinear field configurations can be interpreted as particles with a topologically conserved integer charge $N$ identified as their particle number. Nuclei are described as quantised skyrmions \cite{AdkinsNappiWitten1983static} and the nucleon-nucleon interaction as quantised skyrmion-skyrmion dynamics. An effective hamiltonian can then be extracted and compared to the phenomenological models mentioned above. Authors have used the Skyrme model to rederived the Yukawa pion exchange potential \cite{VinhMau:1984jsh} and find the central  \cite{Jackson:1985bn, WAH1992} and spin-orbit potentials \cite{Nyman:1986py, Otofuji:1987wb, Kaelbermann:1995ed, Abada:1996jg}. Generally, one major problem is found in the skyrmion description of the nucleon-nucleon interaction: the spin-orbit force has the wrong sign.

Most of the previous calculations rely on the product approximation and first order perturbation theory. Motivated by an argument based on the geometry of the 2-skyrmion space, the authors of this paper calculated the isoscalar spin-orbit potential using the dipole approximation and second order perturbation theory \cite{Halcrow:2020gbm}. Despite an initial apparent success, where the derived potential matched the phenomenological one, we recently found an error in the calculation. The corrected spin-orbit potential has the wrong sign for physically interesting parameters. However, neither the dipole or product approximation properly account for the geometry of the 2-skyrmion space. Hence our negative result isn't wholly surprising. In this paper, we will re-calculate the nucleon-nucleon interaction in the Skyrme model using the only known approximation which does account for the geometry of the 2-skyrmion space: the instanton approximation.

Instantons are solutions of Yang-Mills theory in $\mathbb{R}^4$. Atiyah and Manton first showed that one could approximate a charge $N$ skyrmion by taking a holonomy of a charge $N$ instanton \cite{AtiyahManton1989}. This method has been used to construct symmetric skyrmions with low charges \cite{LeeseManton1994stable, houghton1999-3skyrme, sutcliffe2004Buckyball} and more elaborate families of configurations \cite{Cork:2021uov}. Revealing why this works, Sutcliffe showed that Yang-Mills theory is equivalent to a Skyrme model coupled to an infinite tower of vector mesons \cite{Sutcliffe:2010et}. The standard Skyrme model is then the first term in this infinite series. Despite these successes, the instanton approximation has barely been used to model nuclear physics in the Skyrme model. Two exceptions are an initial investigation of the skyrmion-skyrmion interaction \cite{Hosaka:1991kh} and a construction of a simplified Deuteron wavefunction \cite{Leese:1994hb}. 

One reason why instantons have not been widely used to model skyrmions is their numerical complexity. To calculate an instanton-generated skyrmion, naively one must solve an ODE (the parallel transport) at each point in three-dimensional space. Then to calculate the skyrmion-skyrmion interaction one must generate a four-dimensional manifold of configurations (which accounts for relative separations and orientations). Overall, we should solve an ODE at each point in a seven-dimensional space: a daunting task. Fortunately, a new method to construct instanton-generated skyrmions has recently been developed \cite{Harland:2022ohz}. Using the ADHM construction, the parallel transport equation can be solved by multiplying projection operators, which can be done at great speed. As such, we have been able to calculate the potential and metric for two interacting skyrmions in the instanton approximation for the first time. This is the first serious comparison of the instanton approximation against the dipole and product approximations, and we find major disagreements.

Our main result is a calculation of the nucleon-nucleon interaction for the Skyrme model based on the dipole, product and instanton approximations. To obtain this we generalised the second order perturbation theory developed in \cite{Harland:2021dkj}. Although this quantisation procedure is cumbersome, the final result is of broad applicability. To facilitate dissemination of this result, we have generated data files of the NN potentials in terms of the classical metric and potential of the 2-skyrmion interaction. These will allow other researchers to quickly calculate the nucleon-nucleon interaction using different versions of the Skyrme model and different approximations to skyrmion dynamics. We hope this stimulates research which closer links Skyrme theory to practical nuclear physics. A guide for how to use the files can be found in appendix \ref{app:using}.

The paper is structured as follows. In section 2 we describe the classical two-skyrmion interaction and its symmetries. We also describe a non-trivial check of a vital sign in the 2-skyrmiom metric. We then carefully describe the dipole, product and instanton approximations in section 3, including a comparison between them. Section 4 describes the quantum calculation and section 5 contains the final results and a comparison to phenomenological models.

\section{Skyrmion-skyrmion dynamics}
\label{sec:2}

\subsection{Dynamics of a single skyrmion}
The Skyrme model describes the interactions of nuclei using an $SU(2)$-valued field $U$.  The dynamics of $U$ is governed by the lagrangian
\begin{equation}
	\int_{\RR^3} \left(-\frac{F_\pi^2}{16\hbar}\tr(L_\mu L^\mu) + \frac{\hbar}{32 e^2}\tr([L_\mu,L_\nu][L^\mu,L^\nu]) - \frac{F_\pi^2m_\pi^2}{8\hbar^3}\tr(1_2-U)\right)\dd^3y,
\end{equation}
in which $F_\pi$ is the pion decay constant, $e$ is a dimensionless constant, $L_\mu = U^{-1}\partial_\mu U$ is the left-invariant current and we parametrise space with $\boldsymbol{y} \in \mathbb{R}^3$. Throughout this paper we work in Skyrme units of energy ($F_\pi/4e$) and length ($2\hbar/eF_\pi$); in these units the lagrangian is
\begin{equation} \label{eq:dimlessLag}
	\int_{\RR^3} \left(-\frac{1}{2}\tr(L_\mu L^\mu) + \frac{1}{16}\tr([L_\mu,L_\nu][L^\mu,L^\nu]) - m^2\tr(1_2-U)\right)\dd^3y,
\end{equation}
with $m=2m_\pi/eF_\pi$.  This separates into kinetic and potential energy:
\begin{equation} \label{eq:TandV}
	\begin{aligned}
		T &= \int_{\RR^3} \left(-\frac{1}{2}\tr(L_0 L_0) - \frac{1}{8}\tr([L_0,L_i][L_0,L_i])\right)\dd^3y\\
		V &= \int_{\RR^3} \left(-\frac{1}{2}\tr(L_i L_i) - \frac{1}{16}\tr([L_i,L_j][L_i,L_j]) + m^2\tr(1_2-U)\right)\dd^3y.
	\end{aligned}
\end{equation}
These are invariant under isorotations $g\in SU(2)$, acting as $U\mapsto gUg^{-1}$.  They are also invariant under rotations $h\in SU(2)$, which in quaternionic notation act as $\bx\mapsto h\bx h^{-1}$, and under the parity transformation, which acts as $U(\bx)\mapsto U^{-1}(-\bx)$.  Skyrmions are energy-minimising static solutions of the resulting field equations with boundary condition $U = 1_2$ at $|\by| = \infty$. Such fields are maps between three-spheres and as such can be classified by a topologically conserved integer $N$, physically identified with the baryon number. In this paper we will consider massless pions by choosing $m=0$.

The 1-skyrmion has spherical symmetry and takes the form
\begin{equation} \label{eq:1skyrmion}
	U_H(\by) = \exp\left( -i f(|\boldsymbol{y}|) \hat{y}_a\sigma_a\right) \, ,
\end{equation}
where $\sigma_a$ are the Pauli matrices, and the profile function $f$ satisfies
\begin{equation}
	\left(|\boldsymbol{y}|^2+2\sin^2f\right)f'' + 2|\boldsymbol{y}| f' + \sin 2f \left( f'^2-1-\frac{\sin^2 f}{|\boldsymbol{y}|^2} \right) = 0
\end{equation}
and the boundary conditions $f(0) = \pi$ and $f \sim C/|\boldsymbol{y}|^2$ as $|\boldsymbol{y}| \to \infty$. The constant $C$ can be calculated numerically: for $m=0$ it is 2.1596.  We can generate a manifold of equal-energy 1-skyrmions, related by translations $\bx\in\mathbb{R}^3$ and isorotations $Q\in SU(2)$. The fields take the form
\begin{equation}\label{eq:1skyrmionmoduli}
	U(\by;\bx, Q) = Q U_H(\by-\bx) Q^{-1} \,. 
\end{equation}
The low-energy dynamics of a 1-skyrmion can be approximated by treating $\bx$ and $Q$ as functions of $t$; substituting the resulting dynamical Skyrme field into eq.\ \eqref{eq:dimlessLag} results in a Lagrangian,
\begin{equation}
	L = \frac{M}{2}|\dot{\bx}|^2 + \frac{\Lambda}{2}|\bomega|^2 - M,
\end{equation}
in which $\bomega$ is defined by $2Q^{-1}\dot{Q}=-\ii \bomega\cdot\bsigma$.  This is a lagrangian for a moving and rotating body; the mass $M$ and moment of inertia $\Lambda$ can be calculated numerically and take the approximate values $\Lambda=106.83$, $M=145.85$.

\subsection{The 2-skyrmion lagrangian and its symmetries} \label{sec:symms}

The focus of this article is the dynamics of two skyrmions.  There are many ways to model the dynamics of two skyrmions, and these will be described in the next section.  In this section we describe the general form of a 2-skyrmion lagrangian, based on physical expectations and the underlying symmetries.

A system of two skyrmions should be described by two position vectors $\bx_1,\bx_2\in\RR^3$ and two orientations $Q_1,Q_2\in SU(2)$.  We choose to work in a centre of mass frame $\bx_1+\bx_2=\boldsymbol{0}$ and introduce the separation vector $\bx=\bx_1-\bx_2$.  The most general 2-skyrmion lagrangian that is quadratic in velocities takes the form,
\footnote{The notation in this lagrangian has changed slightly from our earlier paper \cite{Harland:2021dkj}: we have absorbed $\rho$ into $A,B,C$ and $D$, written $B^{12}_{ij}=B^{21}_{ji}=B_{ij}$ and introduced $B^{11}_{ij},B^{22}_{ij}$.}
\begin{equation}\label{eq:skyrmionlagrangian}
	L = \frac{M}{4}\dot{x}^i\dot{x}^i+\frac{\Lambda}{2}\omega_\alpha^i\omega_\alpha^i + C_{ij}\dot{x}^i\dot{x}^j+A^\alpha_{ij}\dot{x}^i\omega_\alpha^j + \frac{1}{2}B^{\alpha\beta}_{ij}\omega_\alpha^i\omega_\beta^j-2D,
\end{equation}
where $A, B, C$ and $D$ are functions of $\bx,Q_1,Q_2$. The indices $i,j$ run from 1 to 3 and $\alpha,\beta$ run from 1 to 2. The vectors $\bomega_\alpha$ are defined by
\begin{equation}
	Q_\alpha^{-1}\dot{Q}_\alpha = -\frac{\ii}{2}\omega_\alpha^j\sigma_j.
\end{equation}
The kinetic energy part of this lagrangian can be rewritten in terms of a $9\times 9$ matrix $g$ as follows:
\begin{equation}
	T = \frac{1}{2}\begin{pmatrix}\dot{\bx} & \bomega_1 & \bomega_2 \end{pmatrix} g \begin{pmatrix}\dot{\bx} \\ \bomega_1 \\ \bomega_2 \end{pmatrix},\quad g=\begin{pmatrix}\frac{M}{2}I_3+2C & A^1 & A^2 \\ (A^1)^T & \Lambda I_3+B^{11} & B^{12} \\ (A^2)^T & B^{21} & \Lambda I_3+B^{22}\end{pmatrix}.
\end{equation}
This matrix $g$ defines a metric on the configuration space $\RR^3\times SU(2)\times SU(2)$.  At large separations the lagrangian should agree with the lagrangian of two non-interacting 1-skyrmions, and this means that the functions $A,B,C,D$ should decay as $|\bx|\to\infty$.

The lagrangian should be invariant under the action of isorotations and rotations, parametrised by $g,h\in SU(2)$ and acting as follows:
\begin{align}
	(\bx,Q_1,Q_2) &\mapsto (\bx,gQ_1,gQ_2)\label{eq:isorotations}\\
	(\bx,Q_1,Q_2) &\mapsto (h\bx h^{-1},Q_1h^{-1},Q_2h^{-1})\label{eq:rotations}.
\end{align}
It should also be invariant under reversal of parity:
\begin{equation}
	(\bx,Q_1,Q_2) \mapsto (-\bx,Q_1,Q_2)\label{eq:parity}.
\end{equation}
Since the two skyrmions are indistinguishable, it should be invariant under 
\begin{equation}
	(\bx,Q_1,Q_2) \mapsto (-\bx,Q_2,Q_1)\label{eq:relabelling}.
\end{equation}
Finally, it should be invariant under
\begin{equation}\label{eq:signflips}
	\begin{aligned}
		(\bx,Q_1,Q_2) &\mapsto (\bx,-Q_1,Q_2) \\
		(\bx,Q_1,Q_2) &\mapsto (\bx,Q_1,-Q_2),
	\end{aligned}
\end{equation}
because the single skyrmion \eqref{eq:1skyrmionmoduli} satisfies $U(\by;\bx,-Q)=U(\by;\bx,Q)$.

Isorotation invariance \eqref{eq:isorotations} implies that the coefficients $A^\alpha,B^{\alpha\beta},\ldots$ can be written as functions of $\bx$ and $Q:=Q_1^{-1}Q_2$.   Invariance under sign flips \eqref{eq:signflips} means that all coefficient functions are invariant under $Q\mapsto-Q$.  Invariance under parity \eqref{eq:parity} and relabelling \eqref{eq:relabelling} implies that
\begin{equation}\label{constraints1}
	\begin{aligned}
		A^1(\bx,Q^{-1})&=A^2(\bx,Q),\\ 
		B^{11}(\bx,Q^{-1})&=B^{22}(\bx,Q),\\
		B^{12}(\bx,Q^{-1})&=B^{21}(\bx,Q),\\
		C(\bx,Q^{-1})&= C(\bx,Q),\\
		D(\bx,Q^{-1})&= D(\bx,Q). 
	\end{aligned}
\end{equation}

Since the lagrangian is invariant under rotations \eqref{eq:rotations}, the coefficient functions $A,B,C,D$ are fully determined by their values at points $\bx=(0,0,r)$ with $r>0$.  It will prove convenient to represent these functions using an expansion in spherical harmonics of $Q$.  We only use the first two sets of even spherical harmonics, i.e.\ the constant function and the functions $R_{ab}(Q)=\frac12\operatorname{Tr}(\sigma_a Q \sigma_b Q^{-1})$.  Thus we write:
\begin{equation}\label{harmonic expansion}
	\begin{aligned}
		A^\alpha_{ij}((0,0,r),Q)&=A^\alpha_{0;ij}(r)+A^\alpha_{ab;ij}(r)R_{ab}(Q)\\
		B^{\alpha\beta}_{ij}((0,0,r),Q)&=B^{\alpha\beta}_{0;ij}(r)+B^{\alpha\beta}_{ab;ij}(r)R_{ab}(Q)\\
		C_{ij}((0,0,r),Q)&=C_{0;ij}(r)+C_{ab;ij}(r)R_{ab}(Q)\\
		D((0,0,r),Q)&=D_{0}(r)+D_{ab}(r)R_{ab}(Q).
	\end{aligned}
\end{equation}
The functions of $r$ on the right of \eqref{harmonic expansion} will be referred to as Fourier coefficients.  Note that the functions on the left of \eqref{harmonic expansion} are necessarily invariant under $Q\to-Q$, so odd spherical harmonics do not appear in the expansions on the right hand side.  The constraints \eqref{constraints1} can be rewritten as constraints on the Fourier coefficients, using the fact that $R_{ab}(Q^{-1})=R_{ba}(Q)$:
\begin{equation}\label{constraints2}
	\begin{aligned}
		A^2_{0;ij} &= A^1_{0;ij}, & A^2_{ab;ij}&=A^1_{ba;ij}, &
		B^{22}_{0;ij} &= B^{11}_{0;ij}, & B^{22}_{ab;ij}&=B^{11}_{ba;ij}, \\
		B^{21}_{0;ij} &= B^{12}_{0;ij}, & B^{21}_{ab;ij}&=B^{12}_{ba;ij}, &
		C_{ab;ij}&=C_{ba;ij},& D_{ab}&=D_{ba}. & 
	\end{aligned}
\end{equation}

By choosing to work on the positive $x^3$-axis we have broken some of the rotational symmetry of the lagrangian.  However, our lagrangian should still be invariant under the action of the subgroup $O(2)\subset O(3)$ which fixes points on the axis.  This group acts as follows:
\begin{equation}
	\begin{pmatrix}\dot{x}_1\\\dot{x}_2\\\dot{x}_3\\\end{pmatrix} \mapsto \left(\begin{array}{cc|c}R&&0\\&&0\\\hline 0&0&1\end{array}\right)\begin{pmatrix}\dot{x}_1\\\dot{x}_2\\\dot{x}_3\\\end{pmatrix},\quad
	\begin{pmatrix}\omega^\alpha_1\\\omega^\alpha_2\\\omega^\alpha_3\end{pmatrix}\mapsto \det(R)\left(\begin{array}{cc|c}R&&0\\&&0\\\hline 0&0&1\end{array}\right)\begin{pmatrix}\omega^\alpha_1\\\omega^\alpha_2\\\omega^\alpha_3\end{pmatrix}.
\end{equation}
Invariance under this action imposes further constraints on the Fourier coefficients, which we summarise in the next subsection.

\subsection{Constraints on the Fourier coefficients} \label{sec:constraints}

$A^\alpha_{0;ij}$ has just one independent component, which we take to be $A^1_{0;12}$.  All components with at least one lower index equal to `3' vanish.   The other components can be expressed as follows:
\begin{equation*}
	A^1_{0;21}=-A^1_{0;12},\quad
	A^1_{0;11}=A^1_{0;22}=0,\quad
	A^2_{0;ij}=A^1_{0;ij}.
\end{equation*}

$A^1_{ab;ij}$ has nine independent components, which we take to be $A^1_{12;11}$, $A^1_{21;11}$, $A^1_{11;12}$, $A^1_{33;12}$, $A^1_{23;13}$, $A^1_{32;13}$, $A^1_{23;31}$, $A^1_{32;31}$, $A^1_{12;33}$.  All components with an odd number of `3's in their lower indices vanish.  All components whose lower indices are a permutation of $1122$, $1133$ or $2233$ vanish. All components whose four lower indices are all equal vanish. The remaining components can be expressed as follows:
\begin{equation*}
	\begin{aligned}
		&A^1_{22;21}=-A^1_{11;12}, \quad
		A^1_{12;22}=-A^1_{21;11},\quad 
		A^1_{33;21}=-A^1_{33;12},\quad
		A^1_{21;33}=-A^1_{12;33}, \\ 
		&A^1_{31;32}=-A^1_{32;31},\quad
		A^1_{31;23}=-A^1_{32;13},\quad
		A^1_{13;32}=-A^1_{23;31}, \quad
		A^1_{13;23}=-A^1_{23;13},\\ 
		&A^1_{21;22}=-A^1_{12;11}, \quad A^1_{22;12} = -A^1_{11;21}=A^1_{11;12}+A^1_{21;11}+A^1_{12;11},
	\end{aligned}
\end{equation*}
$A^2_{ab;ij}$ has no further components, because $A^2_{ab;ij}=A^1_{ba;ij}$.

$B^{12}_{0;ij}$ has two independent components, which we take to be $B^{12}_{0;11}$ and $B^{12}_{0;33}$.  Any component with exactly one lower index equal to `3' vanishes.  The remaining components can be expressed as follows:
\begin{equation*}
	B^{12}_{0;22}=B^{12}_{0;11},\quad B^{12}_{0;12}=B^{12}_{0;21}=0.
\end{equation*}
$B^{21}_{0;ij}$ has no further components, because $B^{21}_{0;ij}=B^{12}_{0;ij}$.

$B^{12}_{ab;ij}$ satisfies the identity $B^{12}_{ba;ji}=B^{12}_{ab;ij}$ due to \eqref{constraints2} and the fact that the metric is a symmetric matrix.  This tensor has eight independent components, which we take to be $B^{12}_{22;11}$, $B^{12}_{12;12}$, $B^{12}_{21;12}$, $B^{12}_{33;11}$, $B^{12}_{11;33}$, $B^{12}_{13;13}$, $B^{12}_{31;13}$, $B^{12}_{33;33}$.  Any component with an odd number of `3's amongst its lower indices vanishes.  Any component whose lower indices are a permutation of 1112, 1222 or 1233 vanishes.  The remaining components are
%\begin{align*}
%	B^{12}_{12;21}&=B^{12}_{21;12}&
%	B^{12}_{11;11}&=B^{12}_{22;11}+B^{12}_{12;12}+B^{12}_{21;12}\\
%	B^{12}_{21;21}&=B^{12}_{12;12}&
%	B^{12}_{22;22}&=B^{12}_{22;11}+B^{12}_{12;12}+B^{12}_{21;12}\\
%	B^{12}_{11;22}&=B^{12}_{22;11} &&\\
%	B^{12}_{33;22}&=B^{12}_{33;11}&
%	B^{12}_{22;33}&=B^{12}_{11;33}\\
%	B^{12}_{31;31}&=B^{12}_{13;13}&
%	B^{12}_{13;31}&=B^{12}_{31;13}\\
%	B^{12}_{23;23}&=B^{12}_{13;13}&
%	B^{12}_{32;23}&=B^{12}_{31;13}\\
%	B^{12}_{32;32}&=B^{12}_{13;13}&
%	B^{12}_{23;32}&=B^{12}_{31;13}
%\end{align*}
\begin{align*}
	&B^{12}_{12;21}=B^{12}_{21;12}, \quad
	B^{12}_{21;21}=B^{12}_{12;12}, \quad
	B^{12}_{11;11}=B^{12}_{22;22}=B^{12}_{22;11}+B^{12}_{12;12}+B^{12}_{21;12}\\ &
	B^{12}_{11;22}=B^{12}_{22;11}, \quad
	B^{12}_{33;22}=B^{12}_{33;11}, \quad
	B^{12}_{22;33}=B^{12}_{11;33} \\ &
	B^{12}_{31;31}=B^{12}_{23;23}=B^{12}_{32;32}=B^{12}_{13;13}, \quad
	B^{12}_{13;31}=B^{12}_{32;23}=B^{12}_{23;32}=B^{12}_{31;13}.
\end{align*}
$B^{21}_{ab;ij}$ has no further components, because $B^{21}_{ab;ij}=B^{12}_{ba;ij}(=B^{12}_{ab;ji})$.

$B^{11}_{0;ij}$ has two independent components, which we take to be $B^{11}_{0;11}$ and $B^{11}_{0;33}$.  Any component with exactly one lower index equal to `3' vanishes.  The remaining components can be expressed as follows:
\begin{equation*}
	B^{11}_{0;22}=B^{11}_{0;11},\quad B^{11}_{0;12}=B^{11}_{0;21}=0.
\end{equation*}
$B^{22}_{0;ij}$ has no further components, because $B^{22}_{0;ij}=B^{11}_{0;ij}$.

$B^{11}_{ab;ij}$ satisfies the identity $B^{11}_{ab;ji}=B^{11}_{ab;ij}$.  It has seven independent components, which we take to be $B^{11}_{11;11}$, $B^{11}_{12;12}$, $B^{11}_{33;11}$, $B^{11}_{11;33}$, $B^{11}_{13;13}$, $B^{11}_{31;31}$, $B^{11}_{33;33}$.  Any component with an odd number of `3's amongst its lower indices vanishes.  Any component whose lower indices are a permutation of 1112, 1222 or 1233 vanishes.  The remaining components are
%\begin{align*}
%	B^{11}_{21;12}&=B^{11}_{12;12}&
%	B^{11}_{22;22}&=B^{11}_{11;11}\\
%	B^{11}_{12;21}&=B^{11}_{12;12}&
%	B^{11}_{22;11}&=B^{11}_{11;11}-2B^{11}_{12;12}\\
%	B^{11}_{21;21}&=B^{11}_{12;12}&
%	B^{11}_{11;22}&=B^{11}_{11;11}-2B^{11}_{12;12}\\
%	B^{11}_{33;22}&=B^{11}_{33;11}&
%	B^{11}_{22;33}&=B^{11}_{11;33}\\
%	B^{11}_{13;31}&=B^{11}_{13;13}&
%	B^{11}_{31;13}&=B^{11}_{31;31}\\
%	B^{11}_{23;23}&=B^{11}_{13;13}&
%	B^{11}_{32;23}&=B^{11}_{31;31}\\
%	B^{11}_{23;32}&=B^{11}_{13;13}&
%	B^{11}_{32;32}&=B^{11}_{31;31}.
%\end{align*}
\begin{align*}
	&B^{11}_{22;22} = B^{11}_{11;11}, \quad B^{11}_{21;21} = B^{11}_{21;12}=B^{11}_{12;21}=B^{11}_{12;12}, \quad
	B^{11}_{33;22}=B^{11}_{33;11},\\
	&B^{11}_{22;33}=B^{11}_{11;33},\quad
	B^{11}_{23;32}=B^{11}_{23;23}=B^{11}_{13;31}=B^{11}_{13;13}, \\
	&B^{11}_{32;23}=B^{11}_{32;32}=B^{11}_{31;13}=B^{11}_{31;31}, \quad 
	B^{11}_{11;22}=B^{11}_{22;11} =B^{11}_{11;11}-2B^{11}_{12;12}  .
\end{align*}
$B^{22}_{ab;ij}$ has no further components, because $B^{22}_{ab;ij}=B^{11}_{ba;ij}$.

The constraints on $C_{ab;ij}$ are the same as those on $B^{11}_{ab;ij}$, simply with $B^{11}$ replaced by $C$.

Finally, $D$ has three independent components, which we take to be $D_0$, $D_{11}$ and $D_{33}$.  Then $D_{22}=D_{11}$ and the remaining components are zero.

In summary, we have expressed the potential and metric using a truncated Fourier expansion (i.e.\ an expansion in spherical harmonics, or Wigner D-matrices).  Due to symmetry, the Fourier coefficients can be expressed as functions on a four-dimensional subspace of the nine-dimensional configuration space.  Altogether our expansion has 460 Fourier coefficients, but the symmetries mean that only 41 are independent. In practice, we calculate all the coefficients and use the symmetry relationships as a check on our numerical method. We describe the calculation of the Fourier coefficients in detail in the next section.

\subsection{Expected behaviour of the coefficient functions}
\label{sec:2.4}

It is known that the minimal-energy 2-skyrmion has axial symmetry.  This fact has implications for the metric $g$ and the Fourier coefficients \eqref{harmonic expansion}, which we explore in this section.

We consider a pair of skyrmions in the attractive channel.  This means that their relative orientation is a 180 degree rotation about an axis orthogonal to their axis of separation.  We take the axis of separation to be the $x^3$-axis and the axis of rotation to be the $x^1$-axis.  Then, up to isorotation, the separation vector and orientations are
\begin{equation}
\label{attractive channel}
	(\bx,Q_1,Q_2) = (r\mathbf{k},1,\mathbf{i})
\end{equation}
for some $r>0$.

It is known that the skyrmions in this channel attract one another, and that at a certain separation $r_0$ they merge to form the toroidal energy-minimising 2-skyrmion.  This 2-skyrmion has axial symmetry about the $x^1$-axis (not the $x^3$-axis, as might naively be expected).  More precisely, the 2-skyrmion is invariant under a combination of a rotation $e^{\theta\mathbf{i}/2}\in SU(2)$ and isorotation $e^{\theta\mathbf{i}}\in SU(2)$, for any angle $\theta\in[0,2\pi)$.  If we apply the same transformation to a configuration with $r>r_0$ we generate a path in configuration space of the form
\begin{equation}
\label{short path}
\begin{aligned}
	(\bx,Q_1,Q_2) &= (re^{\theta\mathbf{i}/2}\mathbf{k}e^{-\theta\mathbf{i}/2},e^{\theta\mathbf{i}}e^{-\theta\mathbf{i}/2},e^{\theta\mathbf{i}}\mathbf{i}e^{-\theta\mathbf{i}/2})\\
	& = (re^{\theta\mathbf{i}/2}\mathbf{k}e^{-\theta\mathbf{i}/2},e^{\theta\mathbf{i}/2},\mathbf{i}e^{\theta\mathbf{i}/2}),\,\theta\in[0,2\pi)
\end{aligned}
\end{equation}
This path describes a pair of skyrmions orbiting the $x^3$-axis.  Each skyrmion is also spinning about an axis through its centre, and the directions of spinning and orbiting are opposite.
When $r=r_0$ this path has length zero, so for $r>r_0$ we expect it to be relatively short.  In particular, it should be shorter than the following similar-looking path:
\begin{equation}
\label{long path}
	(\bx,Q_1,Q_2) = (re^{-\theta\mathbf{i}/2}\mathbf{k}e^{\theta\mathbf{i}/2},e^{\theta\mathbf{i}/2},\mathbf{i}e^{\theta\mathbf{i}/2}).
\end{equation}
This path, like \eqref{short path}, describes a pair of spinning and orbiting skyrmions, but this time the directions of spinning and orbiting are the same.

The tangent vectors to these two paths are given by $(\pa_\theta\bx,\bomega_1,\bomega_2)$ with $\bomega=2Q^{-1}\pa_\theta Q$.  When $\theta=0$, they are
\begin{equation}
\begin{aligned}
	U &= (-r\mathbf{j},\mathbf{i},\mathbf{i}) \\
	V &= (r\mathbf{j},\mathbf{i},\mathbf{i}).
\end{aligned}
\end{equation}
Our expectation is therefore that
\begin{equation}\label{gVVgUU}
	g(V,V)-g(U,U)>0
\end{equation}
for small $r$.  In fact, $g(V,V)-g(U,U)$ tends to zero as $r\to\infty$, so assuming monotonicity we expect \eqref{gVVgUU} to hold for all $r$.

In terms of the metric components we have that
\begin{equation}
\begin{aligned}
	g(U,U) &= \frac{1}{2}(M+4C_{22})r^2 -2r\sum_{\alpha}A^\alpha_{21} + (2\Lambda^2+\sum_{\alpha,\beta}B^{\alpha\beta}_{11})\\
	g(V,V) &= \frac{1}{2}(M+4C_{22})r^2 +2r\sum_{\alpha}A^\alpha_{21} + (2\Lambda^2+\sum_{\alpha,\beta}B^{\alpha\beta}_{11}).
\end{aligned}
\end{equation}
So the constraint \eqref{gVVgUU} becomes
\begin{equation}
	\sum_\alpha A^\alpha_{21}>0.
\end{equation}
The matrix $R_{ab}(\mathbf{i})$ is diagonal with entries 1,-1,-1, so $A^\alpha_{21}=A^\alpha_{0;21}+A^\alpha_{11;21}-A^\alpha_{22;21}-A^\alpha_{33;21}$.  In terms of the independent components listed in section \ref{sec:constraints}, the constraint is therefore
\begin{equation}\label{test1}
	-A^1_{0;12}-A^1_{21;11}-A^1_{12;11}+A^1_{33;12}>0.
\end{equation}
The equation \eqref{test1} that we have just derived is useful for two reasons.  First, it serves as a useful test on our numerical calculation of the Fourier coefficients for the instanton approximation (as the instanton approximation reproduces the axial symmetry of the energy-minimising 2-skyrmion).  And second, it will give some insight into the sign of the spin-orbit potential.

\section{The dipole, product and instanton approximations}
\label{sec:3}

In this section we consider the dipole, product and instanton approximations to the 2-skyrmion lagrangian. We are primarily concerned with the instanton calculation but as the methods presented here are rather novel, we review the (numerically simpler) dipole and product calculations first.

\subsection{The dipole approximation}

Far from its centre, we can model a single skyrmion as a triplet of dipoles. The metric and potential  can be calculated analytically at large separations $r\gg 1$ by assuming that the skyrmions interact as dipoles in a linear scalar field theory. For more details, see \cite{Harland:2021dkj,Schroers:1993yk}. The tensors $A,B,C,D$ for the dipole lagrangian are\footnote{Note that our calculation includes terms which are not present in \cite{Schroers:1993yk}, as explained in the erratum of \cite{Harland:2021dkj}. Hence the expressions for $A$ and $C$ do not match the calculation from \cite{Schroers:1993yk}}
\begin{align}
	A^1_{ab;ij}&= \rho \epsilon_{ajc}(\tfrac{1}{2}\nabla_{ibc}r-\delta_{ib}\nabla_c\tfrac{1}{r}) \nonumber \\
	B^{12}_{ab;ij}&= \rho \epsilon_{aic}\epsilon_{bjd}\nabla_{cd}r\\
	C_{ab;ij}&= \rho \left( \tfrac{1}{4}\nabla_{abij}r + \tfrac{1}{2}\delta_{ij}\nabla_{ab}\tfrac{1}{r} - \tfrac{3}{8}(\delta_{ai}\nabla_{bj}+\delta_{bj}\nabla_{ai}+\delta_{aj}\nabla_{bi}+\delta_{bi}\nabla_{aj})\tfrac{1}{r}\right)\nonumber\\ 
	D_{ab} &= \rho \nabla_{ab}\tfrac{1}{r}, \nonumber
\end{align}
where $\rho$ is related to the coefficient $C$ in the tail of the 1-skyrmion profile function $f$. In physical units,
\begin{equation}
	\rho = \frac{8\pi \hbar^3C^2}{e^4 F_\pi^2}.
\end{equation} 
We use these expressions to evaluate the independent Fourier coefficients along the positive $x^3$-axis.  Of the 41 independent Fourier coefficients, 23 vanish, and the remaining 18 are
\begin{align*} 
	&-A^1_{23;31}= 2A^1_{11;12} =2A^1_{12;33}=2A^1_{33;12}=-2A^1_{21;11}=-2A^1_{23;13}=-2A^1_{32;31} = \tfrac{\rho}{r^2}\\
	&B^{12}_{11;33}=B^{12}_{33;11}=-B^{12}_{31;13}=\tfrac{\rho}{r}\\
	&4C_{11;11}=2C_{12;12}=8C_{13;13}=8C_{31;31}=\tfrac{2}{3}C_{33;11}=-\tfrac{1}{2}C_{33;33}=\tfrac{\rho}{r^3} \\
	&-\!D_{11} = \tfrac{1}{2}D_{33} = \tfrac{\rho}{r^3} \, .
\end{align*}
% old expressions (replaced on 12/7/22 by DGH):
%\begin{align*} 
%	&A^1_{23;31}= 2A^1_{11;12} =2A^1_{12;33}=2A^1_{33;12}=-2A^1_{21;11}=-2A^1_{23;13} -2A^1_{32;31} = \tfrac{1}{r^2}\\
%	&B^{12}_{11;33}=B^{12}_{33;11}=-B^{12}_{31;13}=\tfrac{1}{r}\\
%	&4C_{11;11}=2C_{12;12}=8C_{13;13}=8C_{31;31}=\tfrac{2}{3}C_{33;11}=\tfrac{1}{2}C_{33;33}=\tfrac{1}{r^3} \\
%	&-\!D_{11} = \tfrac{1}{2}D_{33} = \tfrac{1}{r^3} \, .
%\end{align*}
In particular, all Fourier coefficients with a subscript `0' vanish and all Fourier coefficients associated with $B^{11}$ or $B^{22}$ vanish.

\subsection{Numerical Implementation} \label{sec:nums}

For the product and instanton approximations, we must resort to using a numerical scheme to calculate the metric and potential.

It is convenient to write the $SU(2)$-valued Skyrme field in terms of four constrained scalar fields,
\begin{equation}
	U = \phi_0 + i \phi_i \sigma_i, \quad \text{where} \quad \phi_0^2 + \phi_i\phi_i = 1\, .
\end{equation}
The potential and kinetic energy are then equal to
\begin{align} \label{eq:Vphi}
	V &= \int_{\mathbb{R}^3} \partial_i \phi_a \partial_i \phi_a + \tfrac{1}{2}( \partial_i \phi_a \partial_i \phi_a  \partial_j \phi_b \partial_j \phi_b  - \partial_i \phi_a \partial_i \phi_b  \partial_j \phi_a \partial_j \phi_b  ) \, \text{d}^3y \, ,\\ 
	T &= \int_{\mathbb{R}^3} \dot{\phi}_a\dot{\phi}_a +  \dot{\phi}_a\dot{\phi}_a \partial_i \phi_b \partial_i \phi_b - \dot{\phi}_a \dot{\phi}_b \partial_i \phi_a \partial_i \phi_b  \, \text{d}^3y \, . \label{eq:Tphi}
\end{align}

We approximate the low energy dynamics by only considering field configurations which depend on the separation and orientation parameters of the skyrmions, so that
\begin{equation} \label{eq:collphi}
	\phi = \phi\left(\by; \bx, Q_1, Q_2\right) \, .
\end{equation}
To find the potential energy in this approximation we substitute \eqref{eq:collphi} into \eqref{eq:Vphi}.
To find the metric we promote the coordinates $\bx, Q_1, Q_2$ to time dependent variables and substitute this into \eqref{eq:Tphi}. This gives an expression for the metric tensor $g$ as
\begin{equation} \label{eq:metric}
	g_{AB} = \int_{\mathbb{R}^3} 2  D_A \phi_a D_B \phi_b \left( \delta_{ab} +  \delta_{ab} \partial_i \phi_c \partial_i \phi_c - \partial_i \phi_a \partial_i \phi_b \right) \, \text{d}^3y
\end{equation}
where $A,B=1,\ldots,9$ and $D_A$ is a derivative on the configuration space generated by $\phi\left(\bx, Q_1, Q_2\right)$. Explicitly,
\begin{equation} \label{eq:derivatives}
	\begin{aligned}
		D_A \phi &= \partial_{t} \phi(\by;\bx+t \boldsymbol{e}_A, Q_1, Q_2)\rvert_{t=0} \\
		D_{A+3} \phi &= \partial_{t} \phi(\by;\bx, Q_1 e^{-i t \sigma_A/2}, Q_2)\rvert_{t=0} \\
		D_{A+6} \phi &= \partial_{t} \phi(\by;\bx, Q_1, Q_2 e^{-i t \sigma_A/2}) \rvert_{t=0}\quad \text{for}\quad A=1,2,3 \, .
	\end{aligned}
\end{equation}
These derivatives are generated numerically using a symmetric second order derivative with step size $0.001$. This requires the repeated generation of the Skyrme field $\phi$ at many points in the configuration space. One numerical complexity is that there are two large manifolds: the manifold of configurations, isomorphic to $SU(2)\times SU(2) \times \mathbb{R}^3$, and Euclidean space, $\mathbb{R}^3$. We need to consider each space carefully.

As discussed in the previous section, the potential and metric only depend nontrivially on four coordinates and we can take $(\bx, Q_1, Q_2) = \left((0,0,r), 1, Q\right)$, which we do from now on. The metric and potential are then functions of $(r,Q)$ on $\mathbb{R}^+\times S^3$. Note that although the metric depends on four coordinates, we still need to take all nine derivatives \eqref{eq:derivatives} to calculate $g$. To sample $\mathbb{R}^+$ we take $r\in[1.731,7.731]$\footnote{This range is chosen so that the axial 2-skyrmion, which occurs at $r=1.731$, is included in the calculation.} with lattice spacing $0.1$. To sample $S^3$ we first consider a hypercube in $\mathbb{R}^4$, which has $8$ cubic faces. Each face is sampled evenly by $(u,v,w) \in [-1,1]\times[-1,1]\times[-1,1]$. The cell is then projected onto the sphere using
\begin{equation}
	(u,v,w) \to \frac{(\pm1,u,v,w)}{\sqrt{1+u^2+v^2+w^2}}
\end{equation}
and other permutations. If we sample $p$ points in each $[-1,1]$ interval we will sample $8p^3$ points on $S^3$ in total, which grows quickly.  In testing we sampled 8, 64 and 216 points, finding that 64 points was a good compromise between computation time and accuracy. The symmetry $Q\to-Q$ means that we actually only need to sample half these points.

At each point $(r,Q)$ on the space of configurations we generate a stencil of Skyrme fields, used to find the derivatives \eqref{eq:derivatives}. Each Skyrme field will be concentrated at $\pm r/2$, the positions of the skyrmions, and have polynomially decaying tails. To account for these two facts we introduce new spatial coordinates $\tilde{y}_a \in [-1,1]$. These are taken from \cite{Leese:1994hb}\footnote{There is an error in the definition of $\tilde{y}_3$ in this paper, as it is not continuous. From context, we believe that \eqref{eq:ycoord} is what the authors used.} and depend on the separation $r$. Explicitly

\begin{align} \label{eq:ycoord}
	y_a &=  \frac{\tilde{y}_a}{1-\tilde{y}_a^2} \quad \text{for} \quad a=1,2 \nonumber \\
	y_3 &= \begin{cases}
		r/2 + \frac{2\tilde{y}_3-1}{8(\tilde{y}_3-1)^2} & \text{if } \tilde{y}_3 \geq 1/2\\
		\tilde{y}_3\left(2r(1-\tilde{y}_3) - (1-2\tilde{y}_3) \right)& 1/2 > \tilde{y}_3 \geq 0 
	\end{cases}
\end{align}
with $y_3$ continued so that it is odd. This transformation ensures that most lattice points are concentrated at $y_3 = \pm r/2$, that the transformation is continuously differentiable at $\tilde{y}_3=\pm1/2$ and that the polynomial tails of the metric densities can be accurately captured. We can then use $\phi(\tilde{\by})$ to calculate the potential $V(r,Q)$ \eqref{eq:Vphi} and metric $g_{AB}(r,Q)$ \eqref{eq:metric}.

Once the metric is calculated, the Fourier coefficients \eqref{harmonic expansion}, each of which is a function of $r$, can be computed using simple integration over the 3-sphere. We calculate these three times, on evenly distributed grids $\tilde{\by} \in [-1,1]^3$ with $80^2\times 160$, $90^2\times 180$ and $100^2\times200$ points. The results are extrapolated to approximate the components on an infinite grid. Extrapolation is often neglected in soliton numerics; it is important here as the metric densities fall off slowly.

Overall, we calculate the Fourier coefficients of $A, B, C$ and $D$ on $\mathbb{R}\times S^3$ at $61 \times 32$ points. Each point requires the generation of 19 configurations to find the field and its derivatives \eqref{eq:derivatives} on the configuration space. Hence, for our reported calculation we generate around $37,000$ Skyrme fields: each on three grid sizes. 

\subsection{Product approximation}

The product approximation was first considered in \cite{Skyrme:1962vh} and builds a 2-skyrmion configuration using the 1-skyrmion \eqref{eq:1skyrmion}, i.e.\ the solution to the static equations of motion with baryon number 1. We use the symmetrised product approximation \cite{Nyman:1987db}, which preserves interchange symmetry \eqref{eq:relabelling} between skyrmions. Here, two 1-skyrmion fields $U_1$ and $U_2$ are combined into a field:
\begin{align}
	U(\by;\bx,Q_1,Q_2) &= \frac{1}{2\mathcal{N}}\left(U_1U_2 + U_2U_1\right) \\
	U_1(\by) &= Q_1U_H\left(\by - \bx/2\right)Q_1^{-1} \\ 
	U_2(\by) &= Q_2 U_H\left(\by + \bx/2\right)Q_2^{-1}\, ,
\end{align}
where $\mathcal{N}$ is a normalisation factor which ensures that $U\in SU(2)$. In $\phi=(\phi^0,\boldsymbol{\phi})$ notation the approximation is
\begin{equation}
	\phi = \frac{1}{\mathcal{N}}\left( \phi_1^0\phi_2^0 - \boldsymbol{\phi}_1 \cdot \boldsymbol{\phi}_2, \phi_1^0 \boldsymbol{\phi}_2 + \phi_2^0\boldsymbol{\phi}_1 \right) \, .
\end{equation}
The paper \cite{Schroers:1993yk} shows that the dipole approximation agrees at large separations with a slightly more complicated variant of the product approximation, called the relativised product approximation.  We do not consider the relativised product approximation here, because we are interested in the product approximation mainly as a consistency check on our calculations, and the level of agreement between the ordinary product ansatz and the dipole approximation proved sufficient for these purposes.

\begin{figure}[h] %[!tph]
	\begin{center}
		\includegraphics[width=0.8\textwidth]{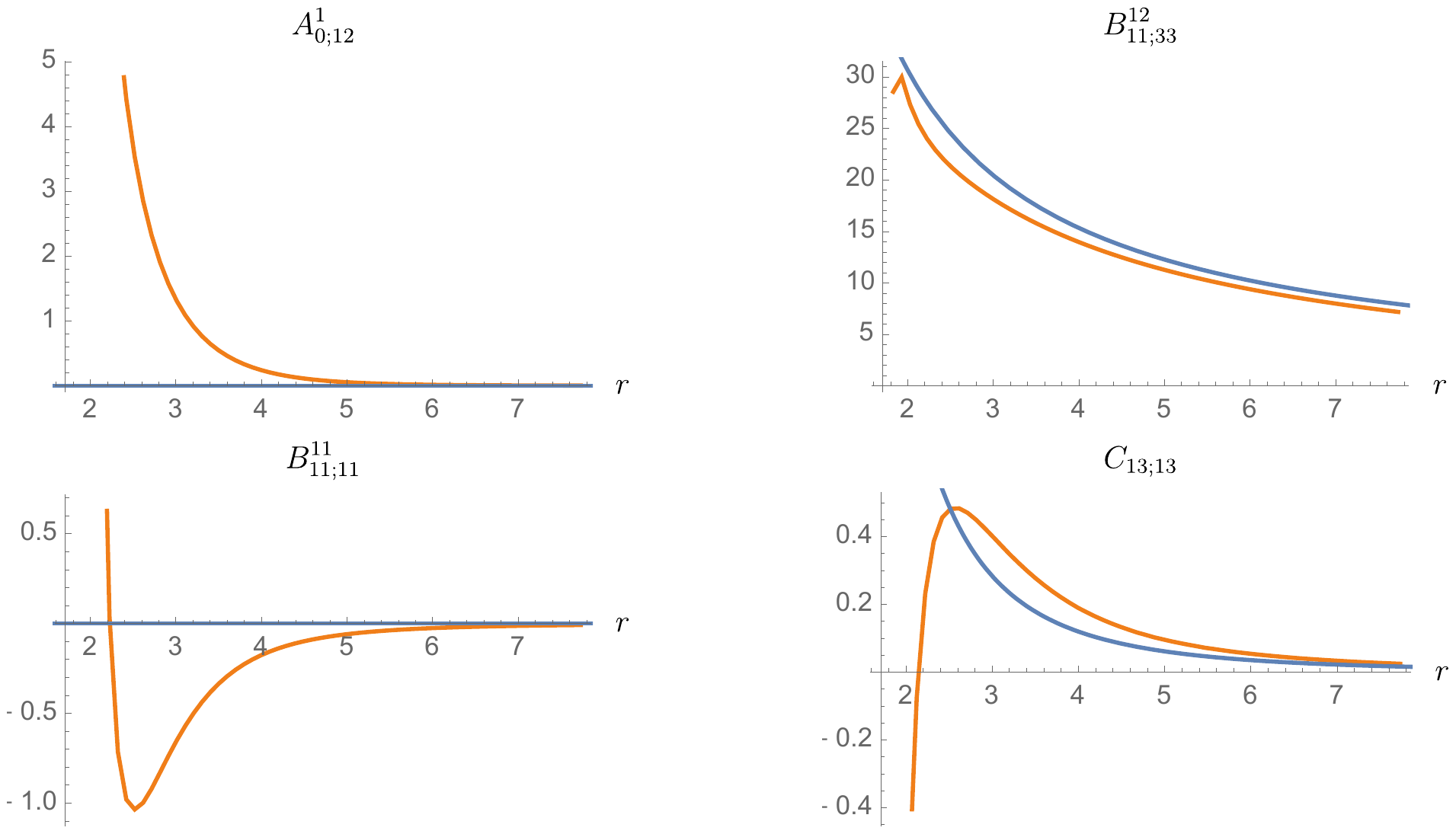}
		\caption{Plots of $A^1_{0;12}$, $B^{12}_{11;33}$, $B^{11}_{11;11}$ and $C_{13;13}$ for the dipole (blue) and product (orange) approximations. We plot the Fourier coefficients as functions of separation $r$ in Skyrme units.}
		\label{fig:prodexamples}
	\end{center}
\end{figure}

We use the product approximation to generate the Fourier coefficients \eqref{harmonic expansion} using the numerical scheme described above. The results match the dipole approximation for large $r$, as expected. We plot four comparisons in Figure \ref{fig:prodexamples}. These demonstrate the possible outcomes when comparing the product and dipole approximations. On the left, we see Fourier coefficients which vanish in the dipole approximation but do not vanish for the product approximation; in these cases, the coefficients obtained from the product approximation decay faster than the leading dipole contribution for that metric term. For instance, the leading dipole contribution to $A$ is $r^{-2}$ and the Fourier coefficients which vanish in the dipole approximation all decay at least as fast as $r^{-3}$. There are many Fourier coefficients that vanish (or are constant) in the dipole approximation, including all coefficients with a subscript ``0''. On the right side of Figure \ref{fig:prodexamples} we see some coefficients which are non-zero in both approximations. For these coefficients, there is always good agreement at large $r$. For $B^{12}_{11;33}$ there is good agreement even at small $r$. For $C_{13;13}$ the dipole and product approximations diverge near $r=3$.

We plot all 41 independent Fourier coefficients for the dipole and product approximations in appendix B. This is the first time such an extensive comparison has been presented and we find nothing particularly surprising. We believe that the calculation in this paper, finding the metric on a moduli space by numerically calculating derivatives, is the first of its kind in 3-dimensions. Hence these non-trivial checks are important for us to have confidence in the method.

\subsection{Instanton approximation}

Atiyah and Manton first suggested that Yang-Mills instantons could be used to approximate skyrmions \cite{AtiyahManton1989,AtiyahManton1993}. A skyrmion is generated by taking holonomies of an $\mathbb{R}^4$-instanton along lines parallel to one of the four coordinate axes. Naively, this entails solving an ODE at each point in space. But recently, a new technique has been developed to calculate the skyrmion efficiently \cite{Harland:2022ohz,Cork:2021ylu} based on the ADHM construction \cite{ADHM1978construction,CWS}. In it, the Skyrme field at a point $\boldsymbol{y}$ is approximated by
\begin{equation} \label{eq:UfromOm}
	U(\by) = \Omega_{\by}(t_n,t_{n-1})\Omega_{\by}(t_{n-1},t_{n-2})\cdots\Omega_{\by}(t_2,t_{1})
\end{equation}
where $t_1,\ldots,t_n$ discretise one direction in $\mathbb{R}^4$ and $\Omega$ are $2\times2$ complex matrices. To construct the $\Omega$ we need the quaternionic $(N+1)\times N$ matrix
\begin{equation}
	\Delta_{y} = \Delta_{(y_0,\by)} = \begin{pmatrix} L \\ M - \left(y_0 \boldsymbol{1} + y_1 \boldsymbol{i} + y_2 \boldsymbol{j} + y_3 \boldsymbol{k} \right) \end{pmatrix} \, ,
\end{equation}
where $L$ and $M$ are matrices of sizes $1 \times N$ and $N\times N$, said to form the ADHM data. The ADHM constraint says that $\Delta_{y}^\dagger \Delta_{y}$ should be real and invertible for all $y\in\mathbb{R}^4$. If this is satisfied, then the ADHM data can be used to construct an instanton and hence a skyrmion. The integer $N$ is identified with the instanton number and hence the baryon number. In this paper $N$ is always $1$ or $2$.

The kernel of $\Delta_{y}$ has quaternionic dimension 1 and can thus be spanned by a single normalised quaternionic vector $v_{y}$. These vectors are then used to construct the $\Omega$. In the simplest numerical scheme \cite{Cork:2021ylu} the $\Omega$ are given by
\begin{equation} \label{eq:simplenums}
	\Omega_{\by}^{1}(t_{n+1},t_n) = v^\dagger_{(t_{n+1},\by)} v_{(t_n,\by)} \, .
\end{equation}
Recently, higher order methods were developed to increase the efficiency of the numerical method \cite{Harland:2022ohz}. We use the ``order 3'' method which, due to special properties of the group $SU(2)$, actually results in an order 4 approximation to the instanton holonomy. In this approximation we replace \eqref{eq:simplenums} with $\Omega^3$ given by
\begin{align*} 
	\Omega^3_{\by}(t_{n+1},t_n) &= \tfrac{4}{3}\Omega_{\by}^2\left(t_{n+1}, \tfrac{1}{2}(t_{n+1}+t_n)\right)\Omega_{\by}^2\left( \tfrac{1}{2}(t_{n+1}+t_n), t_n\right) - \tfrac{1}{3}\Omega_{\by}^2\left(t_{n+1},t_n\right)\\
	\Omega^2_{\by}(t,t') &=\tfrac{1}{2}\Omega_{\by}^1\left(t,t'\right)+\tfrac{1}{2}\Omega_{\by}^1\left(t',t\right)^{-1} \, .
\end{align*}
The numerically generated $U(\by)$ will not be an $SU(2)$ matrix due to small numerical errors. We simply project the near-$SU(2)$ final result back into the $SU(2)$ group by dividing through with $\sqrt{\det U}$.  This method is simpler than the Runge--Kutta method used in \cite{LeeseManton1994stable, Leese:1994hb} and is insensitive to the choice of basis vectors $v$, other than at the points $t=\pm\infty$.  We fix the latter gauge freedom by choosing $v^\dagger = (\boldsymbol{1},0,\ldots,0)$ at these boundary points. We have used quaternions throughout this discussion, but these can be converted to complex $2\times 2$ matrices in a straightforward way.

In practice, the instanton fields are most complicated near $t=0$. We make a coordinate transformation $t = \tan(\tau-\pi/2)$, so that most points are concentrated at the origin. The parameter $\tau$ is sampled evenly in $[0,\pi]$, at 42 points.

The 1-instanton can be described by quaternionic ADHM data
\begin{equation} \label{eq:1adhm}
\begin{pmatrix}L\\M\end{pmatrix}=	\begin{pmatrix} \lambda Q \\ \bx \end{pmatrix} \, ,
\end{equation}
where $Q \in S^3$ (a unit quaternion), $\lambda\in\mathbb{R}$  and $\bx = x_1\boldsymbol{i} + x_2 \boldsymbol{j} + x_3 \boldsymbol{k}$ describe the orientation, size and position of the instanton and corresponding skyrmion. The ADHM data \eqref{eq:1adhm} give rise to a skyrmion of the form
\begin{equation}
	U_H(\by;\bx, Q) = Q\exp\left( -i f_I(r) \sigma_i\frac{(\by-\bx)_i}{r}\right)Q^{-1}
\end{equation}
where $r = |\by-\bx|$. This is of the same form as \eqref{eq:1skyrmion} but with profile function 
\begin{equation}
	f_I(r) = \pi\left(1-\left(1+\frac{\lambda^2}{r^2}\right)^{-1/2} \right) \sim \frac{\pi\lambda^2}{2r^2} \, .
\end{equation}
The skyrmion depends on $\lambda$, the size of the instanton. Choosing $\lambda=1.45$ minimises the skyrmion mass, giving a 1-skyrmion with mass and moment of inertia
\begin{equation}
	M^I_1 = 147.24\, , \quad \Lambda^I_1 = 141.03 
\end{equation}
and tail $f(r) \sim 3.30/r^2$. Note that the tail of the instanton-generated skyrmion is larger than the true 1-skyrmion, leading to a significantly larger moment of inertia and a larger value of $\rho$ for comparing to the dipole calculation.

Two well-separated instantons with equal size $\lambda$, positions $\pm \bx/2 = \pm(x_1 \boldsymbol{i} + x_2 \boldsymbol{j} + x_3 \boldsymbol{k})/2$ and orientations $Q_1$ and $Q_2$ can be described by the Christ--Weinberg--Stanton ADHM data \cite{CWS}
\begin{equation} \label{eq:thedata}
\begin{pmatrix}L\\M\end{pmatrix}=	\begin{pmatrix} \lambda \,Q_1 & \lambda \, Q_2 \\ \bx/2 & \chi \\ \chi & -\bx/2 \end{pmatrix} \, ,
\end{equation}
where $\chi = \lambda^2 \bx /(2|x|^2)\left(\bar{Q}_2Q_1 - \bar{Q}_1Q_2 \right) $. At large $|\bx|$ the data diagonalises and can be thought of as two separated skyrmions with positions $\pm \boldsymbol{x}/2$.  When $|\bx| = \sqrt{2}\lambda$ and $Q_2=Q_1\boldsymbol{k}$, the instanton gains toroidal symmetry and reproduces the well known toroidal skyrmion. Here, we lose the notion of individual skyrmions: they merge completely into one object. We take $\boldsymbol{x}$ to represent the separation of the skyrmions, but the validity of this identification is unclear when the skyrmions are close together. One could define the separation based on the root-mean-square baryon radius of the configuration. We repeated the calculation using this identification, but the final results did not change significantly.

We use the ADHM data \eqref{eq:thedata} to generate the 2-skyrmion family by choosing $Q_1 = \boldsymbol{1}$, $Q_2 = Q$ and $\bx = r \boldsymbol{k}$. However, the size parameter $\lambda$ means that there is an additional coordinate compared to the product approximation. We suppress this coordinate by choosing $\lambda$ which minimises the Skyrme energy \eqref{eq:Vphi} at each point $(r,Q)$, using a Newton-Raphson minimisation. This process generates the 4-dimensional family needed to calculate the Fourier coefficients \eqref{harmonic expansion} of the metric and potential. We use the same derivatives, coordinates and grids as the product approximation.

It is worth emphasising that our ansatz \eqref{eq:thedata} explores only a 10-dimensional subspace of the 12-dimensional moduli space of centred instantons.  In physical terms, we have assumed that the two instantons have the same size and that they are both located in the hyperplane $y_0=0$.  It may be possible to obtain lower-energy Skyrme fields by allowing the instantons to have different displacements in the $y_0$-direction or different sizes. This could have an impact on the symmetries discussed in section \ref{sec:symms} and, as such, we have not explored this possibility.

Note that the instanton calculation requires a holonomy to be calculated at each grid point. Hence, we take around 170 billion holonomies in the final instanton calculation. An accurate calculation would have taken considerably longer without the numerical advancements made in \cite{Harland:2022ohz}.

In the results from the instanton approximation, the numerically generated Fourier components are not very smooth. The root of the problem seems to be that $\lambda$ varies with $r$ and $Q$. The numerical bumpiness will be a problem later, as we need to calculate the second derivative of some Fourier coefficients. To avoid the problem, we use a polynomial interpolation to fit the functions. To do so, we do trial approximations over $r\in  [2.273,7.731]$ of the form
\begin{equation}
	a + br^{-n} + cr^{-n-1} + dr^{-n-2}
\end{equation}
for $n=1,..,8$. Of the eight fits, the best one (based on an adjusted $R^2$ measure) is then kept and used. Another possible solution to this problem of smoothness could be to choose $\lambda$ at each $r$ to minimise the average energy across the 3-sphere, but we did not explore this.

\subsection{Comparison of approximations}

We plot all the independent Fourier coefficients for the dipole, product and instanton approximations in Figures \ref{fig:allplots1} and \ref{fig:allplots2} in appendix A, and a representative sample in Figure \ref{fig:instexamples}.

We find significant differences between the instanton approximation and the other two approximations. For example, either the signs ($A^1_{0;12}$ Figure \ref{fig:instexamples} bottom-left) or magnitudes ($A^1_{23;13}$ Figure \ref{fig:instexamples} bottom-right) can be vastly different for the product and instanton approximations. In general, the instanton Fourier coefficients decay much slower than the product ones. Fitting the instanton tails to functions, we find that they always decay at least as fast as $r^{-1}$.

Some of the differences can be explained by the fact that the dipole strength coefficient $C$ in the instanton approximation is larger than that of the true skyrmion by a factor of 1.5.  A change of this magnitude would multiply the dipole Fourier coefficients by a factor of $1.5^2\approx 2.3$.  With this adjustment, components such as $D_{11}, D_{33}$ and $C_{33;11}$ (Figure \ref{fig:instexamples} top-right) do roughly match across the three calculations.

Another notable difference is that the zeroth order potential $D_0$ (Figure \ref{fig:instexamples} top-left) has the same shape in both approximations, but is much larger for the instanton. We will see later that this results in the instanton approximation giving a repulsive (rather than attractive) central potential between nucleons. 

Despite a smattering of Fourier coefficients which do match, the majority do not.  This means that the product and instanton approximations make different predictions for the dynamics of two skyrmions. We expect that the product approximation gives a more reliable model of skyrmion dynamics at large separations, because it agrees with the dipole picture in that regime.  However, the product approximation is unlikely to be reliable at shorter separations, because it doesn't capture the toroidal symmetry of the energy-minimising 2-skyrmion.  Therefore, we expect the instanton approximation to be more reliable at shorter separations, at least in the attractive channel.  Clearly, there is a need to investigate in more detail which approximation gives the most reliable picture of classical skyrmion dynamics, but doing so goes beyond the scope of this paper.

%There have not been any numerical experiments which distinguish the instanton and dipole approximations, so it is not clear which models true skyrmion dynamics better. Until now, it was assumed that there were only small difference between the methods, at least at large separations. Our results show that the approximations are totally different. 

\begin{figure}[h] %[!tph]
	\begin{center}
		\includegraphics[width=0.8\textwidth]{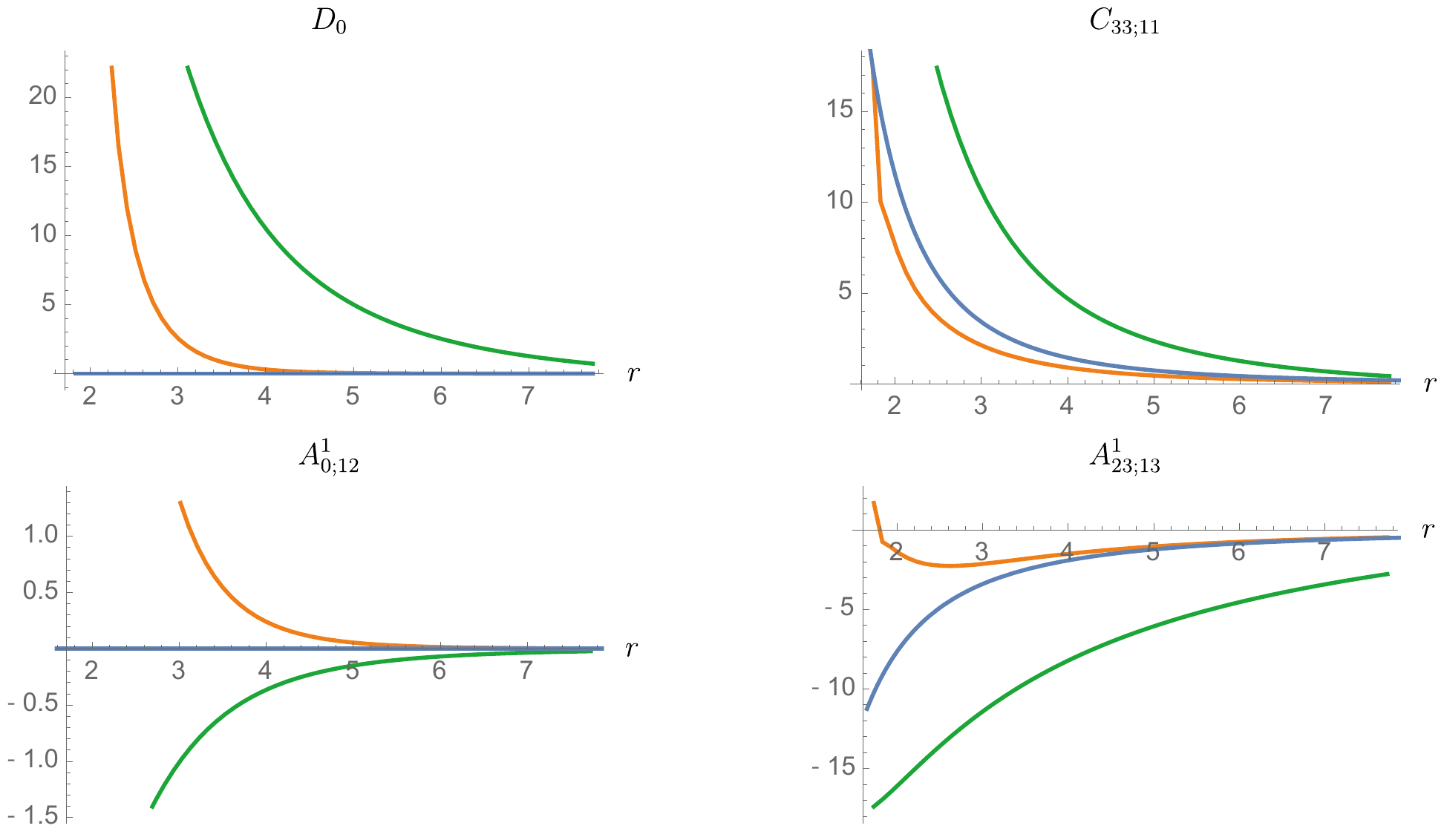}
		\caption{Plots of $D_0$, $C_{33;11}$, $A^1_{12;11}$ and $A^{1}_{23;13}$ for the dipole (blue), product (orange) and instanton (green) approximations. We plot the Fourier coefficients as functions of separation $r$ in Skyrme units. Note that the dipole calculation is calibrated to match the product approximation.}
		\label{fig:instexamples}
	\end{center}
\end{figure}

\section{The nucleon-nucleon potential}
\label{sec:4}

In this section we explain how the nucleon-nucleon potential is calculated from the classical lagrangian for two skyrmions.  This procedure was first used in \cite{Harland:2021dkj}, to which we refer for more details.

The most general nucleon-nucleon potential consistent with known symmetries takes the form:
\begin{align} 
	V_{NN} &= V_C^{IS} + V_{\sigma\sigma}^{IS}\sigsig+V_{12}^{IS}S_{12} + \tfrac{1}{\hbar}V_{LS}^{IS}\Ls \nonumber \\
	&\quad + \left(V_C^{IV} + V_{\sigma\sigma}^{IV}\sigsig+V_{12}^{IV}S_{12} + \tfrac{1}{\hbar}V_{LS}^{IV}\Ls\right)\tautau.
	\label{NN potential}
\end{align}
Here $\sigma_{1i},\sigma_{2i}$ are Pauli matrices corresponding to the spins of two nucleons, and $\sigsig=\sum_i\sigma_{1i}\sigma_{2i}$ is their dot product.  Similarly, $\tautau=\sum_i\tau_{1i}\tau_{2i}$ with $\tau_{1i},\tau_{2i}$ representing the isospins of the two nucleons.  The operator $S_{12}$ is $S_{12}=(3x_{i}x_{j}/r^2-\delta_{ij})\sigma_{1i}\sigma_{2j}$, and $\bL=\bx\times\bP$ is angular momentum.  The eight functions $V_{\ast\ast}^{\ast\ast}$ are assumed to be functions of $r$ only (although in principle they could also depend $|\bP|^2$ and $|\bL|^2$).  The potential \eqref{NN potential} acts on wavefunctions $\psi:\RR^3\to \CC^2\otimes\CC^2\otimes\CC^2\otimes\CC^2$, where the four copies of $\CC^2$ correspond to the spins and isospins of the two nucleons.

The starting point for calculating this from the Skyrme model is the following formula for an effective hamiltonian $H_E$:
\begin{multline}
	\label{EH definition}
	H_E = E_0 + \epsilon H_1^{00} - \epsilon^2\sum_{N>0} \frac{1}{E_N-E_0}H_1^{0N}H_1^{N0} \\
	+\epsilon^3\sum_{M,N\neq0}\frac{1}{(E_N-E_0)(E_M-E_0)}H_1^{0N}H_1^{NM}H_1^{M0} \\
	- \frac{\epsilon^3}{2}\sum_{N>0} \frac{1}{(E_N-E_0)^2}(H_1^{0N}H_1^{N0}H_1^{00}+H_1^{00}H_1^{0N}H_1^{N0}) + O(\epsilon^4).
\end{multline}
In this formula $H=H_0+\epsilon H_1$ is a hamiltonian acting on a large Hilbert space, and $E_0<E_1<\ldots$ are the eigenvalues of $H_0$.  The operator $H_1$ is separated into components $H_1^{NM}$ mapping from the $E_M$-eigenspace to the $E_N$-eigenspace.  When $\epsilon=0$, the restriction of $H$ to the $E_0$-eigenspace is simply $H_E=E_0$, and eq.\ \eqref{EH definition} describes how this effective hamiltonian changes as $\epsilon$ varies away from 0.  Slightly more precisely, \eqref{EH definition} is a perturbative formula for the restriction of $H_0+\epsilon H_1$ to the span of its lowest-energy eigenspaces.

The hamiltonian associated with the lagrangian \eqref{eq:skyrmionlagrangian} can be written 
\begin{equation}\label{eq:skyrmionhamiltonian}
	H=\frac{1}{M}|\bP|^2+\frac{\hbar^2}{2\Lambda}|\bS^1|^2+\frac{\hbar^2}{2\Lambda}|\bS^2|^2+H_I,
\end{equation}
in which $\bP$ is the relative momentum, $\bS^\alpha$ are the spin operators for the two skyrmions, and $H_I$ describes their interaction.  We will apply the perturbative formula to this, identifying $H_0$ with $\frac{\hbar^2}{2\Lambda}\sum_\alpha |\bS^\alpha|^2$ and $\epsilon H_1$ with $\frac{1}{M}|\bP|^2+H_I$.  In doing so, we are assuming that the separation is sufficiently large and the relative motion sufficiently slow that these two terms can be treated as a small perturbation of $H_0$.  Inserting this into \eqref{EH definition} results in
\begin{multline}
	\label{EH}
	H_E = E_0 + \frac{|\bP|^2}{2M}+H_I^{00} - \sum_{N>0}H_I^{0N}\frac{1}{E_N-E_0}H_I^{N0} \\
	%+ \frac{1}{2M}\sum_{N>0}\frac{1}{(E_N-E_0)^2}(H_I^{0N}[|\bP|^2,H_I^{N0}] - [|\bP|^2,H_I^{0N}]H_I^{N0}).
	+ \frac{\ii\hbar}{2M}\sum_{N>0} \frac{1}{(E_N-E_0)^2}\big\{P_i,\, \nabla_iH_I^{0N}H_I^{N0}-H_I^{0N}\nabla_iH_I^{N0}\big\} % This term disagrees with (5.36) Halcrow--Harland 2020!
	\\
	+ \frac{\hbar^2}{M}\sum_{N>0} \frac{1}{(E_N-E_0)^2}\nabla_iH_I^{0N}\nabla_iH_I^{N0}.
\end{multline}
Here we have discarded terms which are cubic in $H_I$ or quadratic in $1/M$ (which is consistent with our assumption that $\frac{1}{M}|\bP|^2+H_I$ is small).  The formula is derived using the fact that $|\bP|^2$ commutes with $H_0$.

The lowest eigenvalues of $H_0=\frac{\hbar^2}{2\Lambda}\sum_\alpha |\bS^\alpha|^2$ are $E_0=\frac{3\hbar^2}{4\Lambda}$, $E_1=\frac{9\hbar^2}{4\Lambda}$ and $E_2=\frac{15\hbar^2}{4\Lambda}$.  The eigenspace associated with $E_0$ is naturally isomorphic to the space of wavefunctions $\psi:\RR^3\to\CC^2\otimes\CC^2\otimes\CC^2\otimes\CC^2$, the Hilbert space for two nucleons.  Moreover, the symmetries \eqref{eq:isorotations}--\eqref{eq:signflips} guarantee that the hamiltonian \eqref{EH} has the same form as the nucleon-nucleon hamiltonian \eqref{NN potential}.

In order to evaluate the effective hamiltonian \eqref{EH definition} one first needs to calculate the classical hamiltonian associated with the 2-skyrmion lagrangian \eqref{eq:skyrmionlagrangian}.  We did this perturbatively, using the identity \cite{Harland:2021dkj}
\begin{align}
	H 
	&= V - \frac{\hbar^2}{2} E_\kappa g_0^{\kappa\lambda} E_\lambda + \frac{\hbar^2}{2} E_\kappa g_0^{\kappa\lambda}\delta g_{\lambda\mu}g_0^{\mu\nu}E_\nu
	- \frac{\hbar^2}{2} E_\kappa g_0^{\kappa\lambda}\delta g_{\lambda\mu}g_0^{\mu\nu}\delta g_{\nu\rho}g_0^{\rho\sigma}E_\sigma \nl
	+ \frac{\hbar^2}{32}\Big[E_{\kappa},\,g_0^{\mu\nu}\delta g_{\mu\nu}\Big]g_0^{\kappa\lambda}\Big[E_\lambda,\,g_0^{\rho\sigma}\delta g_{\rho\sigma}\Big] ,\nl
	{+\frac{\hbar^2}{8}\Big[E_\mu,\,g_0^{\mu\nu}\Big[E_\nu,\,g_0^{\kappa\lambda}\delta g_{\kappa\lambda}-\frac{1}{2}g_0^{\kappa\lambda}\delta g_{\lambda \rho}g_0^{\rho\sigma}\delta g_{\sigma\kappa}\Big]\Big]} \nl
	{-\frac{\hbar^2}{8}\Big[E_\mu,\,g_0^{\mu\nu}\delta g_{\nu\lambda}g_0^{\lambda\kappa}\Big[E_\kappa,\,g_0^{\rho\sigma}\delta g_{\rho\sigma}\Big]\Big]},
\end{align}
in which $g=g_0+\delta g$ and we have neglected terms of order $\delta g^3$.  Substituting the lagrangian \eqref{eq:skyrmionlagrangian} and neglecting terms which are cubic in $A,B,C,D$ or quadratic in $\frac{1}{M}$ results in a hamiltonian of the form \eqref{eq:skyrmionhamiltonian}, with
\begin{align}\label{HI}
	&H_I = 2D
	{
		-\frac{\hbar^2}{8\Lambda^2}[S^\alpha_i,[S^\alpha_i,B^{\beta\beta}_{jj}]]
		-\frac{\hbar^2}{2\Lambda M}[S^\alpha_i,[S^\alpha_i,C_{jj}]]
		-\frac{1}{4M\Lambda}[P_i,[P_i,B^{\alpha\alpha}_{jj}]]
	} \nonumber
	\\ \nonumber
	&-\frac{\hbar^2}{2\Lambda^2}S^\alpha_i B^{\alpha\beta}_{ij} S^\beta_j
	+\frac{\hbar}{M\Lambda}(P^iA^\alpha_{ij}S^\alpha_j+S^\alpha_j A^\alpha_{ij}P_i)
	\\ \nonumber
	&-\frac{\hbar^2}{32\Lambda^3}[S^\gamma_k,B^{\alpha\alpha}_{ii}][S^\gamma_k,B^{\beta\beta}_{jj}]
	-\frac{\hbar^2}{8M\Lambda^2}\left\{\left[S^\gamma_k,B^{\alpha\alpha}_{ii}\right],\left[S^\gamma_k,C_{jj}\right]\right\}
	-\frac{1}{16\Lambda^2M}[P_k,B^{\alpha\alpha}_{ii}][P_k,B^{\beta\beta}_{jj}]
	\\ \nonumber
	&+\frac{\hbar^2}{16\Lambda^3}[S^\gamma_k,[S^\gamma_k,B^{\alpha\beta}_{ij}B^{\alpha\beta}_{ij}]]
		+\frac{\hbar^2}{4\Lambda^2M}[S^\beta_k,[S^\beta_k,A^{\alpha}_{ij}A^{\alpha}_{ij}]]
		+\frac{1}{8\Lambda^2M}[P_k,[P_k,B^{\alpha\beta}_{ij}B^{\alpha\beta}_{ij}]]
	\\ \nonumber
	&+\frac{\hbar^2}{8\Lambda^3}[S^\alpha_i,B^{\alpha\beta}_{ij}[S^\beta_j,B^{\gamma\gamma}_{kk}]]
		+\frac{\hbar^2}{2\Lambda^2M}[S^\alpha_i,B^{\alpha\beta}_{ij}[S^\beta_j,C_{kk}]]
	\\ \nonumber
	&-\frac{\hbar}{4\Lambda^2M}[P_i,A^{\alpha}_{ij}[S^\alpha_j,B^{\beta\beta}_{kk}]]
		-\frac{\hbar}{4\Lambda^2M}[S^\alpha_j,A^{\alpha}_{ij}[P_i,B^{\beta\beta}_{kk}]]
	\\ 
	&+\frac{\hbar^2}{2\Lambda^3}S^\alpha_iB^{\alpha\beta}_{ij}B^{\beta\gamma}_{jk}S^\gamma_k
	+\frac{\hbar^2}{\Lambda^2M}S^\alpha_iA^\alpha_{ji}A^\beta_{jk}S^\beta_k
	-\frac{\hbar}{M\Lambda^2}(P_iA^\alpha_{ij}B^{\alpha\beta}_{jk}S^\beta_k+S^\beta_kA^\alpha_{ij}B^{\alpha\beta}_{jk}P_i).
\end{align}

We have substituted this into the perturbative formula \eqref{EH} and written the resulting expression in the form \eqref{NN potential}.  In order to do this, we made use of the Fourier expansion \eqref{harmonic expansion} of the coefficients $A,B,C,D$, together with the fact that the action of the operators $S^\alpha_i$ and $R_{ab}(Q)$ on the two-skyrmion Hilbert space is known \cite{Harland:2021dkj}.  For example the projections of these operators to the $E_0$-eigenspace are given by 
\begin{equation}
	(S^\alpha_i)^{00} = \frac{1}{2}\sigma_{\alpha i},\quad (R_{ab})^{00} = \frac{1}{9}\sigma_{1a}\sigma_{2b}\tautau.
\end{equation}
The operators $R^{N0}$ vanish when $N>2$, while for $N=1,2$ they describe excitation of one or both nucleons to a delta resonance.

Our calculation also made use of the rotational symmetry of the 2-skyrmion lagrangian in order to simplify terms involving $\bP$.  Suppose, for example, that $\mathcal{Q}$ is a function on the 2-skyrmion configuration space that is invariant under rotations, and that we need to evaluate $[P_i,\mathcal{Q}]$ for $i=1,2,3$.  Along the positive $x^3$-axis, we can write this as
\begin{equation}
	[P_i,\mathcal{Q}] = \delta_{i3}[P_3,\mathcal{Q}]+\frac{1}{r}\epsilon_{3ij}[L_j,\mathcal{Q}]
\end{equation}
The rotational invariance of $\mathcal{Q}$ means that $[L_i+\hbar\sum_\alpha S_i^\alpha,\mathcal{Q}]=0$, so
\begin{equation}\label{eq:identity1}
	[P_i,\mathcal{Q}] =-\ii\delta_{i3}\hbar\frac{d\mathcal{Q}}{dr}-\frac{\hbar}{r}\epsilon_{3ij}\sum_{\alpha=1}^2[S^\alpha_j,\mathcal{Q}].
\end{equation}

Similarly, if $\mathcal{O}_iP_i$ is an operator invariant under rotations then $[L_j+\hbar\sum_\alpha S^\alpha_j,\mathcal{O}_iP_i]=0$, so
\begin{equation}
	\left[L_j+\hbar\sum_\alpha S^\alpha_j,\mathcal{O}_i\right]=\ii\epsilon_{ijk}\mathcal{O}_k.
\end{equation}
It follows that
\begin{align}
	[P_i,\mathcal{O}_i] &= [P_3,\mathcal{O}_3]+\frac{1}{r}\epsilon_{3ij}[L_j,\mathcal{O}_i]\\
	&= -\ii\hbar\left(\frac{d\mathcal{O}_3}{dr}+\frac{2}{r}\mathcal{O}_3\right)-\frac{\hbar}{r}\epsilon_{3ij}\sum_{\alpha=1}^2[S^\alpha_j,\mathcal{O}_i].\label{eq:identity2}
\end{align}
The identities \eqref{eq:identity1} and \eqref{eq:identity2} are useful because they allow us to express the hamiltonian in terms of the Fourier coefficients \eqref{harmonic expansion} and their derivatives in $r$.  They obviate the need to calculate derivatives in all three spacial directions.

To see how these are used in practice consider the term $[P_i,[P_i,B_{jj}^{\alpha\alpha}]]$ on the first line of \eqref{HI}.  This is of the form $[P_i,[P_i,\mathcal{Q}]]$ with $\mathcal{Q}=B_{jj}^{\alpha\alpha}$ invariant under rotations.  So it can be simplified using \eqref{eq:identity1} and \eqref{eq:identity2} to
\begin{equation}
	[P_i,[P_i,B_{jj}^{\alpha\alpha}]] = -\hbar^2\left(\frac{d^2B_{jj}^{\alpha\alpha}}{dr^2}+\frac{2}{r}\frac{dB_{jj}^{\alpha\alpha}}{dr}\right) + \frac{\hbar^2}{r^2}\sum_{i,\beta,\gamma=1}^2[S^\beta_i,[S^\gamma_i,B_{jj}^{\alpha\alpha}]].
\end{equation}
Similarly, the final term on the third line of \eqref{HI} simplifies as follows:
\begin{equation}
	[P_i,B_{jj}^{\alpha\alpha}][P_i,B_{jj}^{\alpha\alpha}] = -\hbar^2\left(\frac{dB_{jj}^{\alpha\alpha}}{dr}\right)^2 + \frac{\hbar^2}{r^2}\sum_{i,\beta,\gamma=1}^2[S^\beta_i,B_{jj}^{\alpha\alpha}][S^\gamma_i,B_{jj}^{\alpha\alpha}].
\end{equation}
Finally, consider the term $P_iA^\alpha_{ij}S^\alpha_j+S^\alpha_jA^\alpha_{ij}P_i$ on the second line of \eqref{HI}.  This can be rewritten as
\begin{align}
	P_iA^\alpha_{ij}S^\alpha_j+S^\alpha_jA^\alpha_{ij}P_i &= \frac12\{P_i,\{A^\alpha_{ij},S^\alpha_j\}\}+\frac12[P_i,[A^\alpha_{ij},S^\alpha_j]]\\
	&= \frac12\{P_i,\{A^\alpha_{ij},S^\alpha_j\}\} - \frac{\ii\hbar}{2}\left[\frac{\partial A^\alpha_{3j}}{\partial r} +\frac{2}{r}A^\alpha_{3j} ,S^\alpha_j\right]\nl
	-\frac{\hbar}{2r}\epsilon_{3ik}\sum_{\beta=1,2}\left[S^\beta_k,\left[A^\alpha_{ij},S^\alpha_j\right]\right].
\end{align}

All of the terms in \eqref{HI} involving $\bP$ can be rewritten in a similar way using \eqref{eq:identity1}, \eqref{eq:identity2}.  Then, using identities presented in \cite{Harland:2021dkj}, they can be substituted into the first line of \eqref{EH} and expressed in terms of the operators $\sigma_{\alpha i}$ etc.  The terms on the second and third lines can similarly be evaluated using identies in \cite{Harland:2021dkj} and the following two identities, both of which follow from \eqref{eq:identity1} and \eqref{eq:identity2}:
\begin{multline}
	\frac{\ii\hbar}{2M}\sum_{N>0}\frac{1}{(E_N-E_0)^2}\{P_i,\nabla_iH_I^{0N}H_I^{N0}-H_I^{0N}\nabla_iH_I^{N0}\} \\
	=\frac{\ii\hbar}{2M}\sum_{N>0}\frac{1}{(E_N-E_0)^2}\{P_3,(H_I')^{0N}H_I^{N0}-H_I^{0N}(H_I')^{N0}\} \\
	+\frac{\hbar}{2Mr}\sum_{N>0}\frac{1}{(E_N-E_0)^2}\sum_{\alpha=1}^{2}\epsilon_{3ij}\big\{P_i,[S_j^\alpha,H^{0N}]H^{N0}-H^{0N}[S_j^\alpha,H^{N0}]\big\}
\end{multline}
\begin{multline}
	\frac{\hbar^2}{M}\sum_{N>0}\frac{1}{(E_N-E_0)^2}\nabla_iH_I^{0N}\nabla_iH_I^{N0}\\
	=\frac{\hbar^2}{M}\sum_{N>0}\frac{1}{(E_N-E_0)^2}(H_I')^{0N}(H_i')^{N0}\\
	-\frac{\hbar^2}{Mr^2}\sum_{N>0}\frac{1}{(E_N-E_0)^2}\sum_{p=1}^2 \sum_{\alpha,\beta}[S^\alpha_p,H_I^{0N}][S^\beta_p,H_I^{N0}].
\end{multline}

We used the Maple software package to carry out the calculation of the hamiltonian \eqref{EH} described above.  After doing so we extracted the eight potentials of \eqref{NN potential} by reading off coefficients.  This resulted in expressions for the eight potentials which were polynomial in the Fourier coefficients, their derivatives, and $\frac{1}{M}$.  As in earlier stages of the calculation, we discarded terms which were either cubic in the 41 independent Fourier coefficients and their derivatives, or quadratic in $\frac{1}{M}$.  The resulting quadratic expressions for $V^{\ast\ast}_{\ast\ast}$ are too lengthy to include in the paper, but we have made them available electronically so that other researchers can investigate the nucleon-nucleon interaction without having to rederive the full quantum hamiltonian. For more details, see appendix \ref{app:using}.  The terms which are linear in the Fourier coefficients are relatively tractable and we write these out in full in appendix \ref{app:linear potentials}.

We carried out two consistency checks on the calculation.  First, we checked that our expression for the effective hamiltonian \eqref{EH} in terms of Fourier coefficients is of the form \eqref{NN potential} (as would be expected due to symmetry).  Second, we checked that our polynomial expressions for the eight potentials are consistent with results obtained in our earlier calculation \cite{Harland:2021dkj} based on the dipole-dipole lagrangian.

\section{Results}
\label{sec:5}

\begin{figure}
	\centering
	\includegraphics[width=0.8\textwidth]{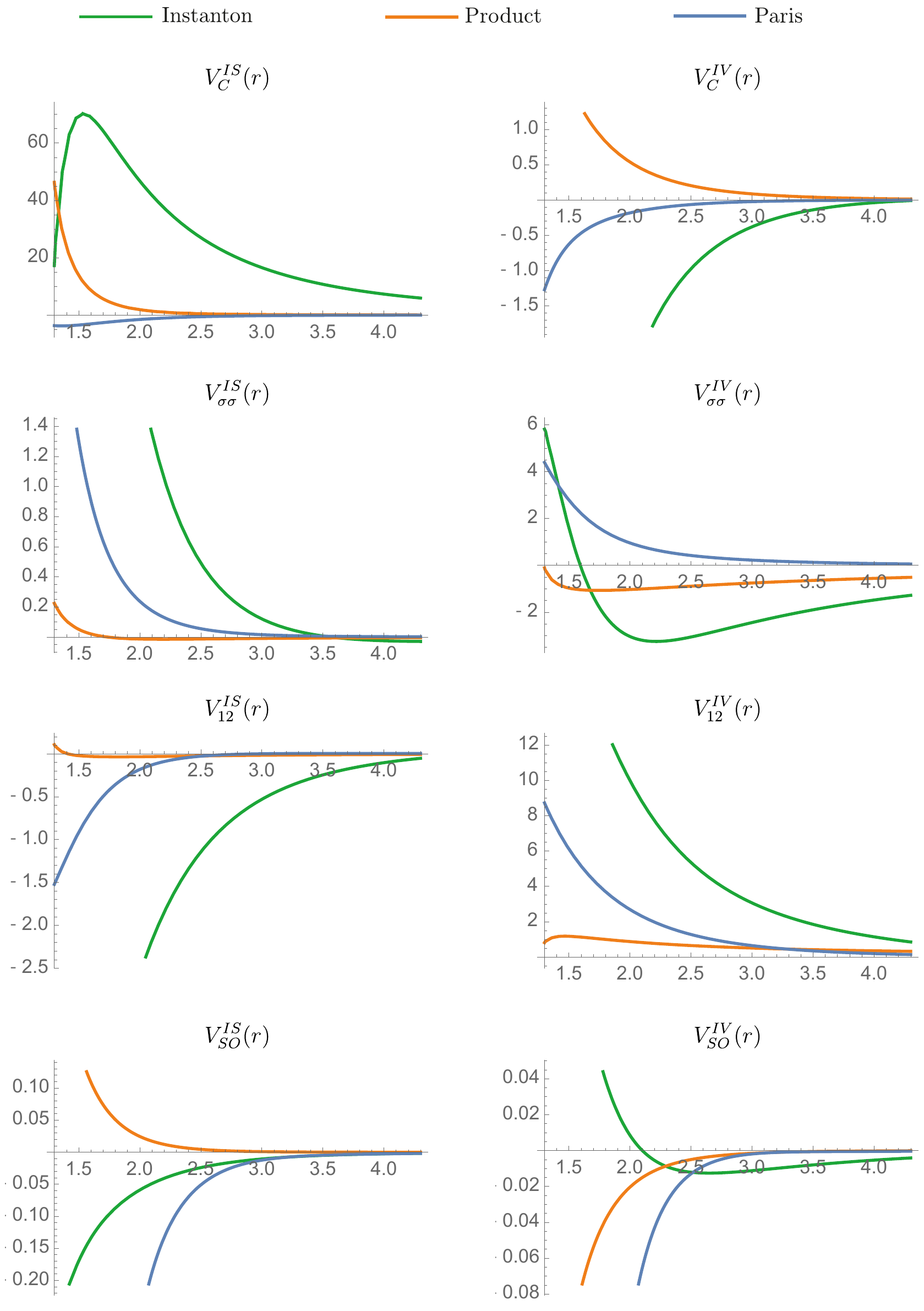}
	\caption{The eight low energy nucleon-nucleon interaction potentials for the product (orange) and instanton (green) approximations, compared to the phenomenological Partis potential (blue). The potentials all have units of energy (MeV) and are plotted as functions of separation (fm).}
	\label{fig:NNpotentials}
\end{figure}
In section \ref{sec:3} we described our calculation of the 41 independent Fourier coefficients of the skyrmion-skymion lagrangian for the product and instanton approximations.  In section \ref{sec:4} we explained how we calculated expressions for the eight components of the nucleon-nucleon potential as polynomials in these Fourier coefficients.  By combining these calculations, we are able to calculate the eight components of the nucleon-nucleon potential in both of these approximations.  The results are plotted in figure \ref{fig:NNpotentials}, alongside the Paris potential.  In order to compare with the Paris potential it is necessary to calibrate the Skyrme model, and we used the calibration of Adkins--Nappi--Witten \cite{AdkinsNappiWitten1983static} (other calibrations did not lead to significantly improved results). We are mainly interested in the results of the instanton approximation; the product approximation is included for comparison.

Taken as a whole, the instanton approximation does not match the Paris potential well.  This is not a surprise.  The instanton approximation assumes that pions are massless, when in reality pions have a mass of around 137MeV.  Our earlier calculation \cite{Harland:2021dkj}, based on the dipole approximation with non-zero pion mass, gave a much better fit to the Paris potential.  Inclusion of a non-zero pion mass is particularly important for the sigma-sigma isovector potential $V_{\sigma\sigma}^{IV}$, because (at least within the dipole approximation) this potential vanishes at first order in perturbation theory when pions are massless, but is non-vanishing at first order when pions are massive.  So we are not concerned that the instanton-generated $V_{\sigma\sigma}^{IV}$ has the opposite sign to the corresponding component of the Paris potential; inclusion of a pion mass should correct this, as it did in \cite{Harland:2021dkj}.

It is more surprising that both the instanton- and product-generated central isoscalar potentials are repulsive, rather than attractive.  In contrast, the dipole approximation with massive pions results in an attractive central isoscalar potential, similar to the Paris model \cite{Harland:2021dkj}.  The product and dipole approximations agree at large separations, so in the case of the product approximation one expects to obtain an attractive central isoscalar potential if the pion mass is set to a realistic value, rather than 0.  Similarly, the repulsive central potential in the instanton approximation could be due in part to the fact that pions are massless in this approximation.

The repulsive central isoscalar potential $V_C^{IS}$ is also likely a consequence of our restricted ansatz \eqref{eq:thedata} for the instantons.  The dominant contribution to $V_C^{IS}$ comes from the Fourier coefficient $D_0$ (see appendix \ref{app:linear potentials}).  As was noted in section \ref{sec:3}, this Fourier coefficient is a positive, decreasing function of $r$, so leads to a repulsive potential.  Second order corrections reduce, but do not overcome, the repulsion seen at first order.  The coefficient $D_0$ is the average over all relative orientations of the interaction energy of two skyrmions.  This interaction energy is negative in the attractive channel but can be positive away from the attractive channel.  Recall that our ansatz \eqref{eq:thedata} did not explore the full moduli space of instantons.  It is likely that, by introducing additional parameters and optimising these to minimise energy, one could obtain a lower average interaction energy and hence a less repulsive $V_C^{IS}$.

The instanton approximation does much better when it comes to the spin-orbit potentials, and in particular the isoscalar spin-orbit potential.  We recall that the dipole approximation, taken to second order in perturbation theory, failed to produce  spin-orbit potentials with the correct sign \cite{Harland:2021dkj}.  In comparison, the instanton approximation gives an isoscalar spin-orbit potential with sign and magnitude comparable with the Paris potential, and an isovector potential with the correct sign at large separations.

We argue that this success is not an accident but is instead a consequence of the fact that the instanton approximation reproduces the axially-symmetric energy-minimising 2-skyrmion.  We recall from section \ref{sec:2} that the existence of an axially-symmetric 2-skyrmion leads to the prediction that
\begin{equation}
	\label{test1revisited}
	-A^1_{0;12}-A^1_{21;11}-A^1_{12;11}+A^1_{33;12}>0.
\end{equation}
On the other hand, to first order in perturbation theory, the spin-orbit potentials are given by
\begin{equation}
	\label{spinorbitlinear}
	V_{LS}^{IS}=\frac{\hbar^{2} A^1_{0;12}}{r \Lambda M},\quad
	V_{LS}^{IV}=\frac{\hbar^{2}}{9 r \Lambda M}(2A^1_{12;11}+A^1_{21;11}+A^1_{32;13}+A^1_{11;12}).
\end{equation}
The simplest way to satisfy the constraint \eqref{test1revisited} is when all four terms on the left hand side are positive.  If this is the case then $A^1_{0;12}$ will be negative and the isoscalar spin-orbit potential will be negative to leading order, consistent with the Paris potential.  Similarly, $A^1_{12;11}$ and $A^1_{21;11}$ will be negative and the isovector spin-orbit potential will likely be negative.  The minimal energy two-skyrmion is axially symmetric for both massless and massive pions. Thus any reasonable approximation to the dynamics of two skyrmions which includes the axial two-skyrmion will likely give correct signs for both spin-orbit potentials, even when the pion mass is switched on.

%We should note that the dipole approximation also satisfies the constraint \eqref{test1revisited}, despite the fact that it doesn't include the axially-symmetric 2-skyrmion.  However, in the dipole approximation $A^1_{0;12}$ (and hence the leading order isoscalar spin-orbit potential) is exactly zero.

It is interesting to consider how this analysis of the spin-orbit force applies to the dipole approximation.  In the dipole approximation, a skyrmion is modelled as a triple dipole with vector-valued charge distribution $\brho_i=CR_i\nabla\delta^3(\bx-\bx_i)$, in which $C>0$ is a positive constant, $R_i$ is a $3\times3$ orthogonal matrix and $\bx_i\in\mathbb{R}^3$.  For a pair of skyrmions in the attractive channel \eqref{attractive channel} we have $\bx_1,\bx_2=\pm r\mathbf{k}/2$ and $R_1,R_2=\mathrm{diag}(\pm 1,\pm 1,1)$.  In particular, when $r=0$ the charge distribution is $\brho = \brho_1+\brho_2=2C\partial_3\delta^3(\bx)$.  This charge distribution is invariant under the action of $SO(2)\times SO(2)$ given by
\begin{equation}
\brho(\bx) \mapsto
\begin{pmatrix}\cos\theta&-\sin\theta&0\\
\sin\theta&\cos\theta&0\\
0&0&1\end{pmatrix}
\brho\left(
\begin{pmatrix}\cos\phi&-\sin\phi&0\\
\sin\phi&\cos\phi&0\\
0&0&1\end{pmatrix}\bx\right).
\end{equation}
This group contains the $SO(2)$ symmetry group of the energy-minimising two-skyrmion as a subgroup.  Therefore the path \eqref{short path} is expected to be short for small values of $r$, for the reasons given in section \ref{sec:2.4}.  However, due to the enhanced $SO(2)\times SO(2)$ symmetry, the path \eqref{long path} is also expected to be short, for similar reasons.  Thus the argument in section \ref{sec:2.4} does not apply to the dipole approximation, and there is no reason to expect negative isoscalar or isovector spin-orbit potentials at first order in perturbation theory.  In fact, the spin-orbit potentials are exactly zero at first order in the dipole approximation; the potentials calculated in \cite{Harland:2021dkj} appear only at second order.  This observation supports our earlier argument that the symmetries of the energy-minimising two-skyrmion explain the signs of the spin-orbit potentials.

%The isovector spin-orbit potential also has the correct sign in the instanton approximation, at least at large separations.  This good agreement is reversed at short separations; it is unclear whether that is due to a breakdown in perturbation theory, or to the limitations of the instanton approximation described above.

\section{Conclusions}

We have calculated the eight components of the nucleon-nucleon potential, starting from an approximation to the dynamics of skyrmions based on instantons.  The two spin-orbit potentials are negative, in agreement with the Paris potential and most other models of nuclear physics.  On the other hand, mixed results were obtained for the remaining six potentials.  In contrast, our earlier calculation \cite{Harland:2021dkj} based on the dipole approximation to skyrmion dynamics agreed with the Paris model for the six potentials that do not involve orbital angular momentum, but gave the wrong sign for the two spin-orbit potentials.

We have argued that the successful results for the spin-orbit potential presented here are due to the symmetry of the energy-minimising two-skyrmion, which is accurately modelled in the instanton approximation but not in the dipole approximation.  The failure of the instanton approximation for the other potentials might be explained by the fact that the pion mass parameter is set to zero in this model, and also by our restricted choice of ansatz for the instantons.  To accurately model the nuclear potential and scattering amplitudes using skyrmions, one would need an approximation to instanton dynamics that reproduces the symmetry of the energy-minimising two-skyrmion and also includes a tuneable pion mass parameter.  Unfortunately, no such approximation is known at present.

Calculating the nucleon-nucleon potential entailed calculating a metric and potential on a space of two-skyrmions generated from instantons.  Surprisingly, these disagreed strongly with two other well-known approximations, the product and dipole approximations, even at large separations.  This raises the question of which approximation is a more reliable guide to the dynamics of skyrmions with massless pions.  On the one hand, the dipole approximation is grounded in well-established physical principles, and for comparable systems involving solitons (such as monopoles) even has the status of a mathematical theorem \cite{stuart}.  On the other hand, the instanton approximation has proven to be very reliable when applied in other situations, such as the study of static energy-minimisers \cite{AtiyahManton1989,LeeseManton1994stable, houghton1999-3skyrme, sutcliffe2004Buckyball,Cork:2021uov}.

The calculation of the potential also involved a calculation in perturbation theory, following a method developed in \cite{Harland:2021dkj}.  This calculation is laborious.  In the present article, we have done the calculation in full generality, expressing our answers in terms of the Fourier coefficients that enter the two-skyrmion lagrangian.  We have made these expressions available for others to use.  This means that a nucleon-nucleon potential can be calculated from any other approximation to two-skyrmion dynamics simply by calculating the Fourier coefficients, and without having to re-do the calculation in perturbation theory.  The files can also be used to study variants of the Skyrme model. An interesting extension would be to add vector mesons, which can be naturally included within the instanton approximation \cite{Sutcliffe:2010et}.

\section*{Acknowledgments}

CJH was supported by the University of Leeds as an academic development fellow and is supported by the Carl Trygger Foundation through the grant CTS 20:25.

%\clearpage
\appendix

\section{All Fourier components for the dipole, product and instanton approximations }
\label{sec:all}

\begin{figure}[H]%[h!] %[!tph]
	\begin{center} 
		\includegraphics[width=\textwidth]{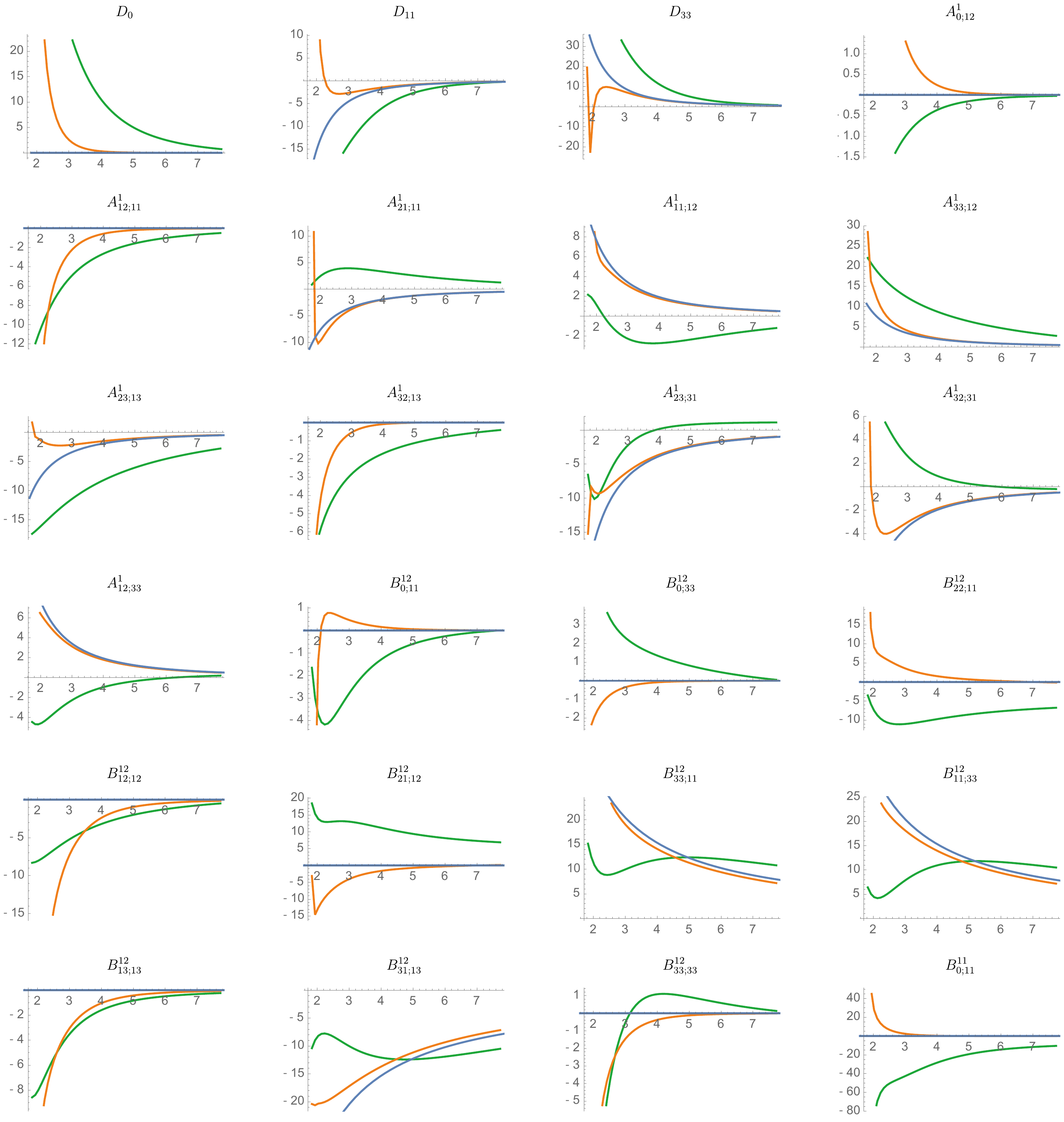}
		\caption{Plots of 24 of the independent Fourier coefficients for the dipole (blue), product (orange) and instanton (green) approximations.}
		\label{fig:allplots1}
	\end{center}
\end{figure}

\begin{figure}[H]%[h!] %[!tph]
	\begin{center}
		\includegraphics[width=\textwidth]{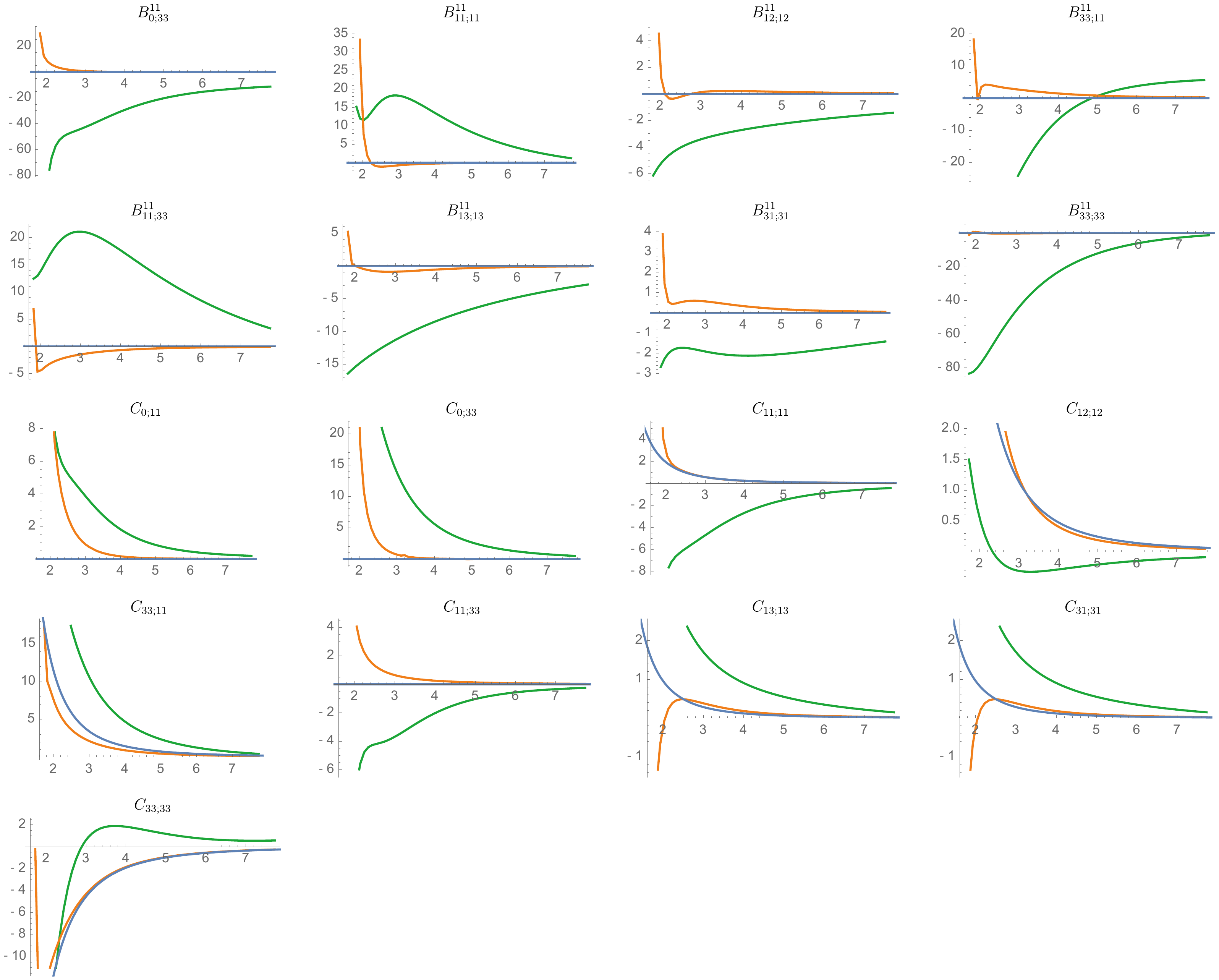}
		\caption{Plots of the remaining 17 independent Fourier coefficients for the dipole (blue), product (orange) and instanton (green) approximations.}
		\label{fig:allplots2}
	\end{center}
\end{figure}

In this appendix, we present all 41 independent Fourier coefficients \eqref{harmonic expansion} for the dipole, product and instanton approximations. These are plotted in Figures \ref{fig:allplots1} and \ref{fig:allplots2}.

\section{Using our results}
\label{app:using}

The supplementary material for this paper includes the full expressions for the eight nuclear potentials that we have calculated. The potentials include all terms which are quadratic in the Fourier coefficients and their derivatives, and linear in $M^{-1}$. The potentials $V_C^{IS}$, $V_{\sigma\sigma}^{IS}$, $V_{12}^{IS}$, $V_{LS}^{IS}$, $V_C^{IV}$, $V_{\sigma\sigma}^{IV}$, $V_{12}^{IV}$ and $V_{LS}^{IV}$ are contained in {\fontfamily{pcr}\selectfont VCIS.txt, VSSIS.txt, V12IS.txt, VSOIS.txt, VCIV.txt, VSSIV.txt, V12IV.txt} and {\fontfamily{pcr}\selectfont VSOIV.txt} respectively. A dictionary between the notation used in this paper and the terms in the text files is displayed in Table \ref{tab:dict}. The text files are formatted so that they can be imported directly into both the Maple and Mathematica software packages.

\begin{table}[h!]
	\begin{center}
		\begin{tabular}{c|c | c | c   }
			\LaTeX  & \texttt{.txt} & \LaTeX  & \texttt{.txt} \\  \hline
			$M$ & \texttt{M}  &$r$ & \texttt{r} \\
			$\Lambda$ & \texttt{La} & $\hbar$  & \texttt{hb} \\ \hline
			\LaTeX & \texttt{.txt} & $\partial_r$ & $\partial_r^2$ \\
			\hline
			$D_0(r)$  & \texttt{D0[r]} & \texttt{dD0[r]}   & \texttt{ddD0[r]}  \\
			$D_{ab}(r)$ & \texttt{D[r,a,b]} & \texttt{dD[r,a,b]} & \texttt{ddD[r,a,b]} \\
			$A_{ij}^p(r)$ & \texttt{A0[r,p,i,j]} & \texttt{dA0[r,p,i,j]} & \texttt{ddA0[r,p,i,j]} \\
			$A_{ab;ij}^p(r)$ & \texttt{A[r,p,a,b,i,j]} & \texttt{dA[r,p,a,b,i,j]} & \texttt{ddA[r,p,a,b,i,j]} \\
			$B^{pq}_{ij}(r)$ & \texttt{B0[r,p,q,i,j]} & \texttt{dB0[r,p,q,i,j]} & \texttt{ddB0[r,p,q,i,j]} \\
			$B^{pq}_{ab;ij}(r)$ & \texttt{B[r,p,q,a,b,i,j]} & \texttt{dB[r,p,q,a,b,i,j]} & \texttt{ddB[r,p,q,a,b,i,j]} \\
			$C_{ij}(r)$ & \texttt{C0[r,i,j]} & \texttt{dC0[r,i,j]} & \texttt{ddC0[r,i,j]} \\ 
			$C_{ab;ij}(r)$ & \texttt{C[r,a,b,i,j]} & \texttt{dC[r,a,b,i,j]} & \texttt{ddC[r,a,b,i,j]} \\
		\end{tabular}
		\caption{Translation from the notation in our paper to the expressions in the appended text files.}
		\label{tab:dict}
	\end{center}
\end{table}

As a typical example, consider the term
\begin{equation}
\frac{\hbar^2}{81\Lambda^2}A^{1}_{2331}(r) \partial_r B^{12}_{2211}(r) \, .
\end{equation}
In the text file, this term becomes
\begin{center}
\texttt{1/81*hb\^{}2/La\^{}2*A[r,1,2,3,3,1]*dB[r,1,2,2,2,1,1]}.
\end{center}

\section{Linear expressions for the potentials}
\label{app:linear potentials}

In these expressions, all terms which are quadratic in the Fourier coefficients have been discarded.
\begin{align*}
	V_{CC}^{IS}&=
	-\frac{\hbar^{2} B^{11}_{0;11}}{2 \Lambda^{2}}-\frac{\hbar^{2} B^{11}_{0;33}}{4 \Lambda^{2}}+2 D_0+\frac{2 \hbar^{2} \partial_r B^{11}_{0;11}}{\Lambda r M}+\frac{\hbar^{2} \partial_r B^{11}_{0;33}}{\Lambda r M}+\frac{\hbar^{2} \partial_r^2 B^{11}_{0;11}}{\Lambda M}+\frac{\hbar^{2} \partial_r^2 B^{11}_{0;33}}{2 \Lambda M}
	\\V_{\sigma\sigma}^{IS}&=
	-\frac{\hbar^{2} B^{12}_{0;11}}{6 \Lambda^{2}}-\frac{\hbar^{2} B^{12}_{0;33}}{12 \Lambda^{2}}
	\\V_{12}^{IS}&=
	\frac{\hbar^{2} B^{12}_{0;11}}{12 \Lambda^{2}}-\frac{\hbar^{2} B^{12}_{0;33}}{12 \Lambda^{2}}
	\\V_{LS}^{IS}&=
	\frac{\hbar^{2} A^1_{0;12}}{r \Lambda M}
	\\V_{CC}^{IV}&=
	-\frac{\hbar^{2} B^{12}_{12;12}}{9 \Lambda^{2}}-\frac{\hbar^{2} B^{12}_{13;13}}{9 \Lambda^{2}}-\frac{\hbar^{2} B^{12}_{33;33}}{36 \Lambda^{2}}-\frac{\hbar^{2} B^{12}_{21;12}}{18 \Lambda^{2}}-\frac{\hbar^{2} B^{12}_{22;11}}{18 \Lambda^{2}}
	\\V_{\sigma\sigma}^{IV}&=
	-\frac{4 \hbar^{2} B^{11}_{11;11}}{27 \Lambda^{2}}+\frac{2 \hbar^{2} B^{11}_{12;12}}{27 \Lambda^{2}}-\frac{\hbar^{2} B^{11}_{33;11}}{18 \Lambda^{2}}-\frac{\hbar^{2} B^{11}_{11;33}}{18 \Lambda^{2}}-\frac{\hbar^{2} B^{11}_{13;13}}{27 \Lambda^{2}}-\frac{\hbar^{2} B^{11}_{31;31}}{27 \Lambda^{2}}-
	\\&\quad
	\frac{5 \hbar^{2} B^{11}_{33;33}}{108 \Lambda^{2}}+\frac{\hbar^{2} B^{12}_{21;12}}{54 \Lambda^{2}}-\frac{\hbar^{2} B^{12}_{22;11}}{54 \Lambda^{2}}+\frac{\hbar^{2} B^{12}_{31;13}}{27 \Lambda^{2}}-\frac{\hbar^{2} B^{12}_{33;11}}{54 \Lambda^{2}}-\frac{\hbar^{2} B^{12}_{11;33}}{54 \Lambda^{2}}+\frac{4 D_{11}}{27}+
	\\&\quad
	\frac{2 D_{33}}{27}+\frac{4 \hbar^{2} A^1_{32;31}}{27 \Lambda r M}-\frac{4 \hbar^{2} A^1_{12;33}}{27 \Lambda r M}-\frac{4 \hbar^{2} A^1_{23;31}}{27 \Lambda r M}+\frac{2 \hbar^{2} \partial_r A^1_{32;31}}{27 \Lambda M}-\frac{2 \hbar^{2} \partial_r A^1_{12;33}}{27 \Lambda M}-
	\\&\quad
	\frac{2 \hbar^{2} \partial_r A^1_{23;31}}{27 \Lambda M}+\frac{4 \hbar^{2} \partial_r B^{11}_{11;11}}{27 \Lambda r M}-\frac{4 \hbar^{2} \partial_r B^{11}_{12;12}}{27 \Lambda r M}+\frac{2 \hbar^{2} \partial_r B^{11}_{33;11}}{27 \Lambda r M}+\frac{2 \hbar^{2} \partial_r B^{11}_{11;33}}{27 \Lambda r M}+\frac{\hbar^{2} \partial_r B^{11}_{33;33}}{27 \Lambda r M}
	\\&\quad
	+\frac{2 \hbar^{2} \partial_r^2 B^{11}_{11;11}}{27 \Lambda M}-\frac{2 \hbar^{2} \partial_r^2 B^{11}_{12;12}}{27 \Lambda M}+\frac{\hbar^{2} \partial_r^2 B^{11}_{33;11}}{27 \Lambda M}+\frac{\hbar^{2} \partial_r^2 B^{11}_{11;33}}{27 \Lambda M}+\frac{\hbar^{2} \partial_r^2 B^{11}_{33;33}}{54 \Lambda M}-\frac{8 \hbar^{2} C_{11;11}}{27 \Lambda M}
	\\&\quad
	+\frac{8 \hbar^{2} C_{12;12}}{27 \Lambda M}-\frac{4 \hbar^{2} C_{33;11}}{27 \Lambda M}-\frac{4 \hbar^{2} C_{11;33}}{27 \Lambda M}-\frac{2 \hbar^{2} C_{33;33}}{27 \Lambda M}
	\\V_{12}^{IV}&=
	\frac{2 \hbar^{2} B^{11}_{11;11}}{27 \Lambda^{2}}-\frac{\hbar^{2} B^{11}_{12;12}}{27 \Lambda^{2}}-\frac{\hbar^{2} B^{11}_{33;11}}{18 \Lambda^{2}}+\frac{\hbar^{2} B^{11}_{11;33}}{36 \Lambda^{2}}-\frac{\hbar^{2} B^{11}_{13;13}}{27 \Lambda^{2}}+\frac{\hbar^{2} B^{11}_{31;31}}{54 \Lambda^{2}}-
	\\&\quad
	\frac{5 \hbar^{2} B^{11}_{33;33}}{108 \Lambda^{2}}+\frac{\hbar^{2} B^{12}_{21;12}}{54 \Lambda^{2}}-\frac{\hbar^{2} B^{12}_{22;11}}{54 \Lambda^{2}}-\frac{\hbar^{2} B^{12}_{31;13}}{54 \Lambda^{2}}+\frac{\hbar^{2} B^{12}_{33;11}}{108 \Lambda^{2}}+\frac{\hbar^{2} B^{12}_{11;33}}{108 \Lambda^{2}}-\frac{2 D_{11}}{27}+
	\\&\quad
	\frac{2 D_{33}}{27}+\frac{\hbar^{2} A^1_{21;11}}{9 \Lambda r M}-\frac{\hbar^{2} A^1_{11;12}}{9 \Lambda r M}-\frac{\hbar^{2} A^1_{23;13}}{9 \Lambda r M}+\frac{\hbar^{2} A^1_{33;12}}{9 \Lambda r M}-\frac{2 \hbar^{2} A^1_{32;31}}{27 \Lambda r M}+\frac{2 \hbar^{2} A^1_{12;33}}{27 \Lambda r M}-
	\\&\quad
	\frac{4 \hbar^{2} A^1_{23;31}}{27 \Lambda r M}-\frac{\hbar^{2} \partial_r A^1_{32;31}}{27 \Lambda M}+\frac{\hbar^{2} \partial_r A^1_{12;33}}{27 \Lambda M}-\frac{2 \hbar^{2} \partial_r A^1_{23;31}}{27 \Lambda M}+\frac{\hbar^{2} B^{11}_{11;11}}{3 \Lambda \,r^{2} M}-\frac{2 \hbar^{2} B^{11}_{12;12}}{3 \Lambda \,r^{2} M}-
	\\&\quad
	\frac{\hbar^{2} B^{11}_{33;11}}{3 \Lambda \,r^{2} M}-\frac{2 \hbar^{2} \partial_r B^{11}_{11;11}}{27 \Lambda r M}+\frac{2 \hbar^{2} \partial_r B^{11}_{12;12}}{27 \Lambda r M}+\frac{2 \hbar^{2} \partial_r B^{11}_{33;11}}{27 \Lambda r M}-\frac{\hbar^{2} \partial_r B^{11}_{11;33}}{27 \Lambda r M}+\frac{\hbar^{2} \partial_r B^{11}_{33;33}}{27 \Lambda r M}-
	\\&\quad
	\frac{\hbar^{2} \partial_r^2 B^{11}_{11;11}}{27 \Lambda M}+\frac{\hbar^{2} \partial_r^2 B^{11}_{12;12}}{27 \Lambda M}+\frac{\hbar^{2} \partial_r^2 B^{11}_{33;11}}{27 \Lambda M}-\frac{\hbar^{2} \partial_r^2 B^{11}_{11;33}}{54 \Lambda M}+\frac{\hbar^{2} \partial_r^2 B^{11}_{33;33}}{54 \Lambda M}+\frac{4 \hbar^{2} C_{11;11}}{27 \Lambda M}-
	\\&\quad
	\frac{4 \hbar^{2} C_{12;12}}{27 \Lambda M}-\frac{4 \hbar^{2} C_{33;11}}{27 \Lambda M}+\frac{2 \hbar^{2} C_{11;33}}{27 \Lambda M}-\frac{2 \hbar^{2} C_{33;33}}{27 \Lambda M}
	\\V_{LS}^{IV}&=
	\frac{2 \hbar^{2} A^1_{12;11}}{9 r \Lambda M}+\frac{\hbar^{2} A^1_{32;13}}{9 r \Lambda M}+\frac{\hbar^{2} A^1_{21;11}}{9 \Lambda r M}+\frac{\hbar^{2} A^1_{11;12}}{9 \Lambda r M}
\end{align*}

\bibliographystyle{unsrt}
\bibliography{refs}
\end{document}